\RequirePackage{xcolor}
\documentclass[journal,twoside,web]{ieeecolor}
\PassOptionsToPackage{usenames,dvipsnames}{xcolor}
\usepackage{generic}
\usepackage{cite}
\usepackage{amsmath,amssymb,amsfonts}
\usepackage{nicefrac}
\usepackage{algorithmic}
\usepackage{graphicx}
\usepackage{algorithm,algorithmic}
\usepackage{hyperref}
\hypersetup{hidelinks=true}
\usepackage{textcomp}
\usepackage{soul}
\usepackage{graphics} 
\usepackage{epsfig} 
\usepackage{mathptmx} 
\usepackage{times} 
\usepackage[acronym]{glossaries}
\usepackage{cleveref}
\usepackage{graphicx}
\usepackage{subcaption}
\usepackage{pstricks} 				
\usepackage{pgfplots} 				
\usepackage{pgfplotstable} 			
\usepackage{tikz}
\usepackage{nicematrix}
\usepackage{commath}
\usepackage{algorithm}
\usepackage{mathtools}
\usepackage{thmtools}
\usepackage{siunitx}
\usepackage{scalerel}
\usepackage{relsize, exscale}
\usepackage{nccmath}
\usepackage{booktabs}
\usepackage{multirow}
\usepackage{bm}
\usepackage{aliascnt}
\def\BibTeX{{\rm B\kern-.05em{\sc i\kern-.025em b}\kern-.08em
    T\kern-.1667em\lower.7ex\hbox{E}\kern-.125emX}}
\newacronym{LTI}{LTI}{linear time-invariant}
\newacronym{SDP}{SDP}{semidefinite program}
\newacronym{MIP}{MIP}{mixed-integer program}
\newacronym{IQC}{IQC}{integral quadratic constraint}
\newacronym{SOS}{SOS}{sum of squares}
\newacronym{ROA}{RoA}{region of attraction}
\newacronym{ReLU}{ReLU}{rectified linear Unit}
\newacronym{IBP}{IBP}{interval bound propagation}
\newacronym{REN}{REN}{recurrent equilibrium network}
\newacronym{RNN}{RNN}{recurrent neural network}
\newacronym{GAS}{GAS}{globally asymptotically stable}
\newacronym{LAS}{LAS}{locally asymptotically stable}
\newacronym[longplural=Linear Matrix Inequalities]{LMI}{LMI}{linear matrix inequality}
\newacronym{LSTM}{LSTM}{long short-term memory}
\newacronym{LQR}{LQR}{linear-quadratic regulator}
\newacronym{MPC}{MPC}{model predictive controller}
\newacronym{DoF}{DoF}{degree of freedom}
\newacronym{RHP}{RHP}{right-half plane}
\newacronym[longplural=equations of motion]{EoM}{EoM}{equation of motion}
\newacronym{CoG}{CoG}{center of gravity}
\newacronym{PRBS}{PRBS}{pseudorandom-binary signal}
\newacronym{MSE}{MSE}{mean-squared error}
\newacronym{NNC}{NNC}{neural-network-based controller}
\newacronym{RMSE}{RMSE}{root mean square error}
\newacronym{MAE}{MAE}{maximum absolute error}
\newcommand{\realsN}[1]{\ensuremath{\mathbb{R}^{#1}}}
\newcommand{\integersN}[1]{\ensuremath{\mathbb{Z}^{#1}}}

\newcommand{\transpose}{^\top}

\DeclareMathAlphabet{\mathcal}{OMS}{cmsy}{m}{n}
\useRomanappendicesfalse
\usepgfplotslibrary{groupplots} 	
\usepgfplotslibrary{colormaps}
\usetikzlibrary{plotmarks}
\usetikzlibrary{calc}
\usetikzlibrary{shapes.geometric}
\usetikzlibrary{arrows.meta,arrows}
\usetikzlibrary{patterns,patterns.meta}
\pgfplotsset{compat=newest}
\pgfplotsset{plot coordinates/math parser=false}
\newlength\figureheight
\newlength\figurewidth
\pgfplotsset{/pgfplots/layers/niceLayers/.define layer set={
		axis background,axis grid,main,axis ticks,axis lines,axis tick labels,axis descriptions,axis foreground
	}{/pgfplots/layers/standard}
}
\pgfplotsset{every axis/.append style={
		set layers=niceLayers,
		tick label style={font=\scriptsize},
		clip marker paths=true,
		line width=1.5pt,
		line cap=round,
		line join=round,
		tick style={semithick, color=black}
}}
\usepgfplotslibrary{external}
\tikzset{external/system call={pdflatex
		 -shell-escape -halt-on-error -interaction=batchmode -jobname "\image" "\texsource"}}
\newtheorem{thm}{Theorem}[section]

\newtheorem{defn}[thm]{Definition}

\crefname{thm}{Theorem}{Theorems}
\crefname{lemma}{Lemma}{Lemmas}
\crefname{prop}{Proposition}{Propositions}
\crefname{cor}{Corollary}{Corollaries}
\crefname{defn}{Definition}{Definitions}
\crefname{conj}{Conjecture}{Conjectures}
\crefname{exmp}{Example}{Examples}
\crefname{rem}{Remark}{Remarks}
\crefname{assume}{Assumption}{Assumptions}
\crefname{equation}{}{} 
\Crefname{equation}{Equation}{Equations} 
\crefname{figure}{Fig.}{Figs.} 
\Crefname{figure}{Fig.}{Figs.} 
\crefname{table}{Table}{Tables} 
\Crefname{table}{Table}{Tables} 
\crefname{subfigure}{Fig.}{Figs.} 
\Crefname{subfigure}{Fig.}{Figs.} 
\crefname{section}{Section}{Sections} 
\crefname{algorithm}{Algorithm}{Algorithms} 
\Crefname{algorithm}{Algorithm}{Algorithms} 
\crefname{appendix}{Appendix}{Appendices}
\Crefname{appendix}{Appendix}{Appendices}  
\DeclareSIUnit{\deg}{deg}
\usepackage{fancyhdr}
\fancypagestyle{firstpage}{
  \fancyhf{}
  
  \fancyhead[C]{\footnotesize This work has been submitted to the IEEE for possible publication. Copyright may be transferred without notice, after which this version may no longer be accessible.}
}
\begin{document}
\title{Synthesis and SOS-based Stability Verification of a Neural-Network-Based Controller for a Two-wheeled Inverted Pendulum}

\author{Alvaro Detailleur, Dalim Wahby, Guillaume Ducard, \IEEEmembership{Senior Member, IEEE},
Christopher Onder
\thanks{A. Detailleur and C. Onder are with the Institute for Dynamic Systems and Control (IDSC), Department of Mechanical and Process Engineering, Swiss Federal Institute of Technology (ETH) Zurich, Leonhardstrasse 21, 8092 Zurich, Switzerland. (E-mail: adetailleur@student.ethz.ch; onder@idsc.mavt.ethz.ch, Telephone: +41 44 632 87 96)}%
\thanks{D. Wahby and G. Ducard are with Universit{\'e} C\^{o}te d`Azur I3S CNRS, 06903 Sophia Antipolis, France. (E-mail: dalim.wahby@univ-cotedazur.fr; guillaume.ducard@univ.cotedazur.fr)}%
}

\maketitle
\thispagestyle{firstpage}
\begin{abstract}
This work newly establishes the feasibility and practical value of a \acrfull{SOS}-based stability verification procedure for applied control problems utilizing \acrfullpl{NNC}. It successfully verifies closed-loop stability properties of a \acrshort{NNC} synthesized using a generalizable procedure to imitate a robust, tube-based \acrfull{MPC} for a two-wheeled inverted pendulum demonstrator system. This is achieved by first developing a state estimator and control-oriented model for the two-wheeled inverted pendulum. Next, this control-oriented model is used to synthesize a baseline \acrfull{LQR} and a robust, tube-based \acrshort{MPC}, which is computationally too demanding for real-time execution on the demonstrator system’s embedded hardware. The generalizable synthesis procedure generates an \acrshort{NNC} imitating the robust, tube-based \acrshort{MPC}. Via an \acrshort{SOS}-based stability verification procedure, a certificate of local asymptotic stability and a relevant inner estimate of the \acrfull{ROA} are obtained for the closed-loop system incorporating this \acrshort{NNC}.
Finally, experimental results on the physical two-wheeled inverted pendulum demonstrate that the \acrshort{NNC} both stabilizes the system, and improves the control performance compared to the baseline \acrshort{LQR} in both regulation and reference-tracking tasks.
\end{abstract}


\section{Introduction}
\label{sec:introduction}
\IEEEPARstart{T}{he} use of neural networks as feedback controllers, so-called \acrfullpl{NNC}, has received significant attention since at least the late 1980s \cite{mybibfile:Hunt1992}. It is widely recognized that they possess several properties that make them uniquely well-suited for such applications. For example, their universal approximation capabilities \cite{mybibfile:Hanin2019} and relatively low computational requirements compared to traditional optimization-based algorithms used in (optimal) control \cite{mybibfile:Gonzalez2024} allow them to represent highly nonlinear and complex control laws, while simultaneously enabling high-frequency evaluation on embedded hardware.

However, the absence of guarantees regarding the closed-loop stability properties of systems controlled by \acrshortpl{NNC} poses a significant obstacle to their adoption in safety-critical applications \cite{mybibfile:Norris2021}. To address these limitations, various optimization-based procedures have been introduced that attempt to certify the stability properties of control loops containing neural networks \cite{mybibfile:Dubach2022,mybibfile:Schwan2023,mybibfile:Richardson2023,mybibfile:Pauli2021, mybibfile:Yin2022, mybibfile:Revay2020, mybibfile:Revay2023,mybibfile:Korda2022,mybibfile:Newton2022}. 

The \acrfull{SOS}-based stability verification procedure utilized in this work \cite{mybibfile:Korda2022,mybibfile:Newton2022} leverages semialgebraic sets to obtain a potentially exact model of the neural network's input-output relationship. This allows the search for a stability certificate to be limited to only those closed-loop trajectories consistent with the neural network’s exact input-output behavior. 

In contrast, frameworks that utilize concepts from robust control theory, such as integral \cite{mybibfile:Richardson2023,mybibfile:Pauli2021, mybibfile:Yin2022} or incremental \cite{mybibfile:Revay2020,mybibfile:Revay2023} quadratic constraints, employ approximate characterizations of neural networks and therefore return stability certificates that simultaneously apply to a (large) set of \acrshortpl{NNC}.

In addition, this procedure makes use of \acrshort{SOS} programming, a subclass of semidefinite programming for which polynomial-time solvers exist. This enables the verification procedure to scale to deep and large networks, potentially consisting of hundreds of neurons \cite{mybibfile:Korda2022}.

However, to the best of the authors' knowledge, practical results demonstrating the application of this procedure to an applied control problem have not yet been reported.

\subsection{Contributions}
\label{sec:Introduction_Contributions}
To the best of our knowledge, this work is the first to apply the aforementioned \acrshort{SOS}-based stability verification procedure for \acrshortpl{NNC} to a real-world problem, namely the control of a two-wheeled inverted pendulum demonstrator system. Using a generalizable synthesis procedure, a \acrshort{NNC} is synthesized to imitate a robust, tube-based \acrfull{MPC} which itself is computationally too expensive to run in real time on the demonstrator system's hardware. This \acrshort{NNC} is proven to possess relevant local closed-loop stability properties using the \acrshort{SOS}-based stability verification procedure. The value of these results is reinforced by experimental results showing that the \acrshort{NNC} outperforms a baseline \acrfull{LQR} in regulation and reference-tracking tasks. Therefore, the main contributions of this work are:

\begin{enumerate}
    \item The presentation of a generalizable synthesis procedure for \acrshortpl{NNC} that imitate robust, tube-based \acrshortpl{MPC}, informed by prior knowledge of the underlying \acrshort{MPC} formulation.
    \item Computation of a certificate of local asymptotic stability and a relevant inner estimate of the \acrfull{ROA} for the two-wheeled inverted pendulum system under a fixed \acrshort{NNC} using an \acrshort{SOS}-based stability verification procedure.
    \item Experimental validation of the locally stable \acrshort{NNC} on a two-wheeled inverted pendulum, including a comparison in which the \acrshort{NNC} outperforms a baseline \acrshort{LQR} in both regulation and reference-tracking tasks.
\end{enumerate}
This represents the first successful application of this \acrshort{SOS}-based stability verification procedure to a real-world control problem and establishes the stability verification procedure's value for the implementation of \acrshortpl{NNC} in safety-critical applications.

This paper aims to document the complete control loop used in the practical experiments. As the (underlying) controllers are synthesized using model-based techniques, 
a description of the state estimator and system model are thus necessary for the contributions presented in this work. Therefore, the paper is organized as follows:
\Cref{sec:SystemAndModelingProcedure} presents the two-wheeled inverted pendulum system, including the hardware and the nonlinear system model. \Cref{sec:StateEstimation} describes the development of a state estimator for the test platform. \Cref{sec:ControlOrientedModel} details the parameter identification procedure and the linearization process used to obtain a control-oriented model suitable for controller synthesis. \Cref{sec:ControllerSynthesis} discusses the synthesis of the baseline \acrshort{LQR} and the robust, tube-based \acrshort{MPC} before presenting the generalizable synthesis procedure for the  \acrshort{NNC} and the modifications applied to the output of all controllers to ensure the assumptions of the control-oriented model are met. \Cref{sec:StabilityVerification} details the stability verification procedure used to analyze the local stability properties of the synthesized \acrshort{NNC}. \Cref{sec:Results} collects and discusses the results that demonstrate the value of the generalizable \acrshort{NNC} synthesis procedure and the \acrshort{SOS}-based stability verification procedure to this applied control problem. These results consist of a qualitative comparison of the three synthesized controllers, an inspection of the proven stability properties of the two-wheeled inverted pendulum controlled by the \acrshort{NNC}, as well as a comparison of the empirical performance of this controller to the aforementioned \acrshort{LQR}. Finally, \cref{sec:Conclusion} presents a conclusion and suggests topics for future research.

\subsection{Notation}
\label{sec:Introduction_Notation}
This work uses the following notational conventions:
\begin{itemize}
    \item In the analysis of discrete-time systems, the plus superscript indicates the successor variable, e.g. $x, \ x^+ \in \realsN{n}$ represent the current and successor state, respectively.
    \item The notation $\{\mathcal{B}\}$ indicates a reference frame in which vector quantities related to rigid-body dynamics can be expressed, e.g. $^{\mathcal{B}}\vec{(\cdot)}$. Derivatives with respect to an inertial frame $\{\mathcal{I}\}$ are denoted $\dot{\vec{(\cdot)}}_{\mathcal{I}}$, rotation matrices transforming vectors from $\{\mathcal{I}\}$ to $\{\mathcal{B}\}$ are denoted ${}^{\mathcal{B}}R_{\mathcal{I}}$, and quantities describing a relative difference between two frames, e.g. the angular velocity of $\{\mathcal{B}\}$ with respect to $\{\mathcal{I}\}$, are (explicitly) denoted $\vec{(\cdot)}_{\nicefrac{\mathcal{B}}{\mathcal{I}}}$. 
    \item All inequalities are defined element-wise. $P \succ 0$, $P \succeq 0$ denote a positive definite and positive semidefinite matrix $P$, respectively.
    \item The set $\{1, \dots, n\}$ is denoted as $[n]$. The subvector of $x \in \realsN{n}$ consisting of the entries indexed by $\mathcal{I} \subseteq [n]$ is denoted $x_{\mathcal{I}}$.  The notation $\mathcal{M}(x, \,n)$ denotes the vector of all unique products of $n$ entries of $x$. 
    \item The identity map is denoted $\textrm{id}$.
    \item The Minkowski sum and Pontryagin difference are denoted by $\oplus$ and $\ominus$, respectively. Elementwise division and multiplication are denoted by $\oslash$ and $\odot$, respectively.
    \item Unless specified otherwise, all norms represent the Euclidean norm.
\end{itemize}

\section{System and Modeling Procedure}
\label{sec:SystemAndModelingProcedure}
\begin{figure}[t]
    \centering
    \def\svgwidth{0.95\columnwidth}
\begingroup%
  \makeatletter%
  \providecommand\color[2][]{%
    \errmessage{(Inkscape) Color is used for the text in Inkscape, but the package 'color.sty' is not loaded}%
    \renewcommand\color[2][]{}%
  }%
  \providecommand\transparent[1]{%
    \errmessage{(Inkscape) Transparency is used (non-zero) for the text in Inkscape, but the package 'transparent.sty' is not loaded}%
    \renewcommand\transparent[1]{}%
  }%
  \providecommand\rotatebox[2]{#2}%
  \newcommand*\fsize{\dimexpr\f@size pt\relax}%
  \newcommand*\lineheight[1]{\fontsize{\fsize}{#1\fsize}\selectfont}%
  \ifx\svgwidth\undefined%
    \setlength{\unitlength}{334.6296758bp}%
    \ifx\svgscale\undefined%
      \relax%
    \else%
      \setlength{\unitlength}{\unitlength * \real{\svgscale}}%
    \fi%
  \else%
    \setlength{\unitlength}{\svgwidth}%
  \fi%
  \global\let\svgwidth\undefined%
  \global\let\svgscale\undefined%
  \makeatother%
  \begin{picture}(1,0.64581693)%
    \lineheight{1}%
    \setlength\tabcolsep{0pt}%
    \put(0,0){\includegraphics[width=\unitlength,page=1]{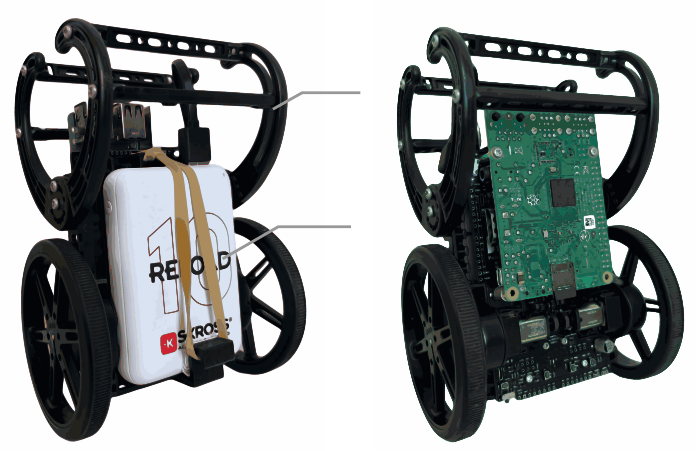}}%
    \put(0.45026491,0.53130121){\makebox(0,0)[lt]{\lineheight{1.25}\smash{\begin{tabular}[t]{l}(i)\end{tabular}}}}%
    \put(0.42600976,0.33956912){\makebox(0,0)[lt]{\lineheight{1.25}\smash{\begin{tabular}[t]{l}(iii)\end{tabular}}}}%
    \put(0,0){\includegraphics[width=\unitlength,page=2]{Sigi_Components_Reduced.pdf}}%
    \put(0.49932561,0.42936887){\makebox(0,0)[lt]{\lineheight{1.25}\smash{\begin{tabular}[t]{l}(ii)\end{tabular}}}}%
    \put(0.50216623,0.2443577){\makebox(0,0)[lt]{\lineheight{1.25}\smash{\begin{tabular}[t]{l}(iv)\end{tabular}}}}%
    \put(0,0){\includegraphics[width=\unitlength,page=3]{Sigi_Components_Reduced.pdf}}%
  \end{picture}%
\endgroup%

    \caption{The Sigi platform, a two-wheeled inverted pendulum, used in this work. This platform consists of (i) the Balboa 32U4 Balancing Robot Kit \cite{mybibfile:PololuBalboa32U4}, (ii) Raspberry Pi 3 Model B+ microprocessor \cite{mybibfile:RaspberryPiProductBrief}, (iii) SKROSS Reload 10 power bank \cite{mybibfile:SkrossReload10Datasheet}, and (iv) Pololu HPCB 6V micro metal gearmotors \cite{mybibfile:PololuMicroMetalGearmotors} with a total approximate 50:1 gear ratio and a quadrature encoder system.}
    \label{fig:SigiPicture}
\end{figure}

In this work an \acrshort{NNC} is synthesized to control a custom, two-wheeled inverted pendulum named Sigi, shown in \cref{fig:SigiPicture}. In this section, a description of the components comprising the Sigi platform and a nonlinear system model used as a basis for controller synthesis are presented.

\subsection{System Description}
\label{sec:SystemAndModelingProcedure_SystemDescription}
The Sigi platform, a two-wheeled inverted pendulum shown in \cref{fig:SigiPicture}, is a custom robotics platform based on the Balboa 32U4 Balancing Robot Kit \cite{mybibfile:PololuBalboa32U4}. This base platform, shown as (i) in \cref{fig:SigiPicture}, consists of the robot chassis, a PCB with an ATmega32U4 AVR microcontroller from Atmel, two DRV8838 Texas Instruments motor drivers, and an STMicroelectronics LSM6DS33 inertial module \cite{mybibfile:PololuLSM6DS33} that consists of a 3-axis accelerometer and a 3-axis gyroscope. Shown as (iv) in \cref{fig:SigiPicture}, the platform is equipped with two Pololu 30:1 HPCB 6V micro metal gearmotors \cite{mybibfile:PololuMicroMetalGearmotors}, each coupled with the provided 41:25 plastic gears and a quadrature encoder system consisting of a magnetic disk and integrated Hall effect sensors. Each quadrature encoder outputs a 16-bit unsigned integer which is incremented $12$ times per rotation. A $\qty[scientific-notation=false,round-precision=1]{5}{\volt}$, $\qty[scientific-notation=false,round-precision=2]{2.4}{\ampere}$ SKROSS Reload 10 power bank \cite{mybibfile:SkrossReload10Datasheet}, shown as (iii) in \cref{fig:SigiPicture}, is used to supply power to the robot. Finally, a Raspberry Pi 3 Model B+ \cite{mybibfile:RaspberryPiProductBrief} microprocessor is used to run the control system algorithms designed in this work.

The Raspberry Pi microprocessor uses an I\textsuperscript{2}C communication protocol to communicate with the aforementioned sensors and actuators. The control system algorithms are designed in MATLAB/Simulink. Using the relevant support packages \cite{mybibfile:MathworksHardwareSupportPackage} and a WiFi connection, these algorithms can be compiled and run wirelessly on the Raspberry Pi microprocessor. Furthermore, the support packages support data acquisition, allowing data to be recorded and analyzed in a desktop MATLAB environment.  

Sigi represents a robotics platform with significant control challenges, due to its nonlinear nature and unstable open-loop dynamics. Furthermore, the platform's small size, $\qty[scientific-notation=false,round-precision=2]{80}{\milli\metre} \times \qty[scientific-notation=false,round-precision=2]{110}{\milli\metre} \times \qty[scientific-notation=false,round-precision=2]{140}{\milli\metre}$ ($\textrm{L} \times \textrm{W} \times \textrm{H})$, and non-minimum phase behavior impose limitations on the fundamentally achievable control performance. In addition, the system exhibits backlash in its gearing system, introducing an additional source of nonlinearity that affects control performance and complicates the development of an accurate mathematical model.
Finally, given the limited power available from the SKROSS Reload 10 power bank, the voltage applied to each of the platform's electric motors cannot exceed $\qty[scientific-notation=false,round-precision=2]{2.0}{\volt}$. These input saturation constraints imply that there exists no globally-stabilizing controller, whilst simultaneously adding further constraints on the chosen control strategy.

\subsection{Nonlinear System Model}
\label{sec:SystemAndModelingProcedure_SystemModel}
\begin{figure*}[t]
    \centering
    \begin{subfigure}{0.375\textwidth} 
        \centering
        \def\svgwidth{\linewidth}
        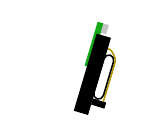
        \caption{}
        \label{subfig:Sigi_FBD_1}
    \end{subfigure}
    \hspace{0.1\textwidth}
    \begin{subfigure}{0.375\textwidth} 
        \centering
        \def\svgwidth{\linewidth}
        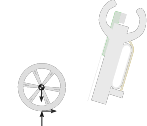
        \caption{}
        \label{subfig:Sigi_FBD_2}
    \end{subfigure}
    \caption{\subref{subfig:Sigi_FBD_1} A schematic diagram of the Sigi platform, showing the inertial and body frames $\{\mathcal{I}\}$, $\{\mathcal{B}\}$, respectively, as well as the definition of the pitch ange $\theta$. \subref{subfig:Sigi_FBD_2} A free-body diagram of the Sigi platform.}
    \centering
    \label{fig:Sigi_FBD}
\end{figure*}

To enable model-based control design, an approximate nonlinear planar model of the Sigi platform is obtained. In this model, the following assumptions about the platform are made:
\begin{itemize}
    \item The robot moves in a plane, over a flat surface perpendicular to gravity. 
    \item The Sigi platform consists of two rigid bodies, the wheel and the pendulum body, attached to each other via a frictionless rotational joint without backlash.
    \item The wheels of the platform satisfy the no-slip condition.
    \item The inductance dynamics of the electric motors are assumed to be instantaneous, allowing the motors to be modeled using an equivalent circuit model consisting of only a resistance and a (back EMF) voltage source. 
    \item All losses present in the motor drivers and voltage regulator are lumped and modeled as a deadband on the applied input voltage to the motor terminals.
\end{itemize}
Under these assumptions, the state of the Sigi platform as observed from an inertial reference frame $\{\mathcal{I}\}$ with origin $\textrm{O}$ can be described by a four-dimensional state vector, $x = [x_w,\, \dot{x}_w,\, \theta,\, \dot{\theta}]\transpose$. As shown in \cref{subfig:Sigi_FBD_1}, $x_w$, $\dot{x}_w$ represents the translational position and velocity of the platform, respectively. With the introduction of a non-inertial body frame $\{\mathcal{B}\}$ attached to the pendulum body at its center of gravity, $\textrm{P}$, the platform's pitch and pitch rate, $\theta$, $\dot{\theta}$, respectively, are also formally defined. 

Following a multibody analysis of the two rigid bodies assumed to comprise the Sigi platform, shown in \cref{subfig:Sigi_FBD_2}, and including all relevant kinematic constraints that follow from the aforementioned assumptions, the \acrfullpl{EoM} for the Sigi platform are given by 
\begin{align}
    \label{eq:SigiEoM_theta}
    & \begin{aligned}
        \ddot\theta & = \frac{1}{d_1} \Big( \Big(2\Big(m_w+\frac{J_w}{r_w^2}\Big)+m_p\Big)\big(m_p l_c g \sin(\theta) - T\big) \\
        & \ \quad - m_p^2 l_c^2 \sin(\theta) \cos(\theta)\dot\theta^2 - m_p l_c \cos(\theta) \frac{T}{r_w} \Big),
    \end{aligned} \\
    \label{eq:SigiEoM_x}
    & \begin{aligned}        
        \ddot x_w & = \frac{1}{d_1}\Big(-m_p^2 l_c^2 \cos(\theta) \sin(\theta) g + T m_p l_c \cos(\theta) \\
        & \ \quad  + \big(J_p + m_p l_c^2\big) m_p l_c \sin(\theta) \dot\theta^2 + \big(J_p + m_p l_c^2\big)\frac{T}{r_w}\Big),
    \end{aligned} \\ 
    \intertext{where}
    \label{eq:SigiEoM_d}
    & \begin{aligned}
       & d_1 = \big(J_p + m_p l_c^2\big)\Big(2\Big(m_w+\frac{J_w}{r_w^2}\Big)+m_p\Big) - m_p^2 l_c^2 \cos^2(\theta),
    \end{aligned}
\end{align}
and the definitions of all parameters are shown in \cref{fig:Sigi_FBD,tab:SigiModelParameters}. 

The total torque $T$ exerted on each of the rigid bodies in the free-body analysis is equal to the sum of the torques produced by the left motor $T_L$ and the right motor $T_R$. Following the assumptions regarding the motors' behavior and the kinematic relation between the pitch angle $\theta$ and the translational position $x_w$, the relationship between the applied input voltage $u$ and the total torque $T$ is given by
\begin{align}
    \label{eq:MotorTorque}
    T &= 2i_{\textrm{gb}} \bigg(\frac{K}{R}\textrm{db}_{u_0}(u) + \frac{K^2}{R}i_{\textrm{gb}}\dot\theta - \frac{K^2}{R}i_{\textrm{gb}}\frac{\dot{x}_w}{r_w}\bigg), 
\end{align}
with the deadband on the input voltage defined as
\begin{align}
    \label{eq:MotorDeadband}
    \textrm{db}_{u_0}(u) &= 
    \begin{cases}
        0                           & \text{if}\; \abs{u} \leq u_0, \\
        \textrm{sign}(u) (|u| - u_0)  & \text{if}\; \abs{u} > u_0.
    \end{cases}
\end{align}
Thus, the full \acrshortpl{EoM} of the Sigi platform in continuous time are given by 
\crefrange{eq:SigiEoM_theta}{eq:MotorDeadband}.

\begin{table}[t]
\centering
\begin{tabular}{lccc}
\hline
\textbf{Quantity}                                                                               & \textbf{Symbol}   & \multicolumn{1}{c}{\textbf{Value}}    & \textbf{Unit}                    \\ \hline
Radius of wheel                                                                                 & $r_w$             & \phantom{00}\num[scientific-notation=false,round-precision=1]{0.04}\phantom{$0e-5$}   & $\si{\metre}$                    \\
Distance between wheels                                                                         & $d$               & \phantom{00}\num[scientific-notation=false,round-precision=1]{0.1}\phantom{$00e-5$}                                  & $\si{\metre}$                    \\
Mass of (single) wheel                                                                          & $m_w$             & \phantom{00}\num[scientific-notation=false,round-precision=1]{0.02}\phantom{$0e-5$}                                & $\si{\kilogram}$                 \\
\begin{tabular}[c]{@{}l@{}}Mass moment of inertia of \\ \ \ \ single wheel around ${}^{\mathcal{I}}\hat{e}_y$ axis\end{tabular} & $J_w$       & \phantom{00}\!\!\num[scientific-notation=true,round-precision=3]{2.2490e-05}\phantom{$0$}                           & $\si{\kilogram\metre\squared}$   \\
Mass of pendulum body                                                                           & $m_p$             & \phantom{00}\num[scientific-notation=false,round-precision=3]{0.368}\phantom{$e-5$}                                  & $\si{\kilogram}$                 \\
\begin{tabular}[c]{@{}l@{}}Mass moment of inertia of\\  \ \ \ pendulum body around ${}^{\mathcal{I}}\hat{e}_y$ axis\end{tabular} & $J_p$      & \phantom{00}\!\!\num[scientific-notation=true,round-precision=3]{3.7600e-04}\phantom{$0$}                              & $\si{\kilogram\meter\squared}$   \\
Pendulum body CoG offset                                                                        & $l_c$             & \phantom{00}\num[scientific-notation=false,round-precision=1]{0.01}\phantom{$0e-5$}                                & $\si{\metre}$                    \\
Gearbox ratio                                                                                   & $i_{\textrm{gb}}$ & \phantom{0}\num[scientific-notation=false,round-precision=4]{49.86}\phantom{$0e-5$}                             & $1$                              \\
Electric motor constant                                                                         & $K$               & \phantom{00}\!\!\num[scientific-notation=true,round-precision=3]{1.4500e-03}\phantom{$0$}                            & $\si{\newton\meter\per\ampere}$  \\
Electric motor resistance                                                                       & $R$               & \phantom{00}\num[scientific-notation=false,round-precision=3]{8.8200e+00}\phantom{$0e-5$}                            & $\si{\ohm}$                      \\
Gravitational acceleration                                                                      & $g$               & \phantom{00}\num[scientific-notation=false,round-precision=3]{9.81}\phantom{$0e-5$}                                   & $\si{\metre\per\second\squared}$ \\ \hline
\end{tabular}
\caption{An overview of the parameters of the Sigi model}
\label{tab:SigiModelParameters}
\end{table}

\section{State Estimation}
\label{sec:StateEstimation}
For the synthesized controllers to effectively control the Sigi platform, a state estimator with sufficient accuracy must be designed. To this end, sensor data 
coming from the 3-axis accelerometer, ${}^{\mathcal{B}}\vec{a}_{\,\nicefrac{\textrm{IMU}}{\textrm{O}}}^{\,\textrm{acc}}$, 3-axis gyroscope, ${}^{\mathcal{B}}\vec{\omega}_{\,\nicefrac{\mathcal{B}}{\mathcal{I}}}^{\,\textrm{gyr}}$, and the unwrapped left and right motor shaft encoder counts, $N^{\textrm{enc}}_{\textrm{L}}$ and $N^{\textrm{enc}}_{\textrm{R}}$, respectively, are used to estimate the attitude of the body frame $\{\mathcal{B}\}$ with respect to the inertial frame $\{\mathcal{I}\}$, as well as the translational position and velocity, $x_w$, $\dot{x}_w$, respectively.

\subsection{Attitude Estimation}
\label{sec:StateEstimation_Attitude}
The attitude of the body frame $\{\mathcal{B}\}$ with respect to the inertial frame $\{\mathcal{I}\}$ is described using yaw-pitch-roll Euler angles. The roll, pitch and yaw angles are denoted as $[\phi, \, \theta, \, \psi]\transpose$, respectively. Estimates of these Euler angles, $[\hat\phi, \, \hat\theta, \, \hat\psi]\transpose$, are obtained by fusing estimates from accelerometer and wheel encoder data $[\hat\phi_{\textrm{acc}}, \, \hat\theta_{\textrm{acc}}, \, \hat\psi_{\textrm{enc}}]\transpose$, with estimates of the Euler angle rates obtained from gyroscope data $[\hat{\dot\phi}_{\textrm{gyr}}, \, \hat{\dot\theta}_{\textrm{gyr}}, \, \hat{\dot\psi}_{\textrm{gyr}}]\transpose$ in a complimentary filter \cite{mybibfile:Hua2014}.

From rigid-body dynamics the relationship between the absolute and relative acceleration of the accelerometer is given by
\begin{equation}
    \begin{aligned}
       \big({}^{\mathcal{B}}\ddot{\vec{r}}_{\,\nicefrac{\textrm{IMU}}{\textrm{O}}}\big)_{\mathcal{I}} & = \ddot{x}_w {}^{\mathcal{B}}R_{\mathcal{I}} {}^{\mathcal{I}}\hat{e}_x + 
       {}^{\mathcal{B}}\dot{\vec{\omega}}_{\,\nicefrac{\mathcal{B}}{{\mathcal{I}}}} \times {}^{\mathcal{B}}\vec{r}_{\,\nicefrac{\textrm{IMU}}{{\textrm{P}}}} \\ & \ \quad
       + {}^{\mathcal{B}}\vec{\omega}_{\,\nicefrac{\mathcal{B}}{{\mathcal{I}}}} \times \big( {}^{\mathcal{B}}\vec{\omega}_{\,\nicefrac{\mathcal{B}}{{\mathcal{I}}}} \times {}^{\mathcal{B}}\vec{r}_{\,\nicefrac{\textrm{IMU}}{{\textrm{P}}}}\big).
   \end{aligned}
\end{equation}
Noting that a DC acceleration opposing the gravity vector is measured by the accelerometer, i.e. ${}^{\mathcal{B}}\vec{a}_{\,\nicefrac{\textrm{IMU}}{\textrm{O}}}^{\,\textrm{acc}} = ({}^{\mathcal{B}}\ddot{\vec{r}}_{\,\nicefrac{\textrm{IMU}}{\textrm{O}}})_{\mathcal{I}} - {}^{\mathcal{B}}\vec{g}$, and assuming that $\ddot{x}_w \ll 1$, an estimate of the gravity vector expressed in the body frame is obtained via
\begin{gather}
    \begin{aligned} 
    \label{eq:BodyFrameGravityVectorEstimate}
        {}^{\mathcal{B}}\vec{\hat{g}} & = -{}^{\mathcal{B}}\vec{a}_{\,\nicefrac{\textrm{IMU}}{\textrm{O}}}^{\,\textrm{acc}} + {}^{\mathcal{B}}\vec{\omega}_{\,\nicefrac{\mathcal{B}}{\mathcal{I}}} \times \big( {}^{\mathcal{B}}\vec{\omega}_{\,\nicefrac{\mathcal{B}}{\mathcal{I}}} \times {}^{\mathcal{B}}\vec{r}_{\,\nicefrac{\textrm{IMU}}{{\textrm{P}}}} \big) \\ & \ \quad + G_{\textrm{BP}}^{\textrm{att}}(s) {}^{\mathcal{B}}\vec{\omega}_{\,\nicefrac{\mathcal{B}}{\mathcal{I}}} \times {}^{\mathcal{B}}\vec{r}_{\,\nicefrac{\textrm{IMU}}{{\textrm{P}}}},
    \end{aligned}
    \intertext{where $G^{\textrm{att}}_{\textrm{BP}}(s)$ represents the band-pass filter}
    G^{\textrm{att}}_{\textrm{BP}}(s) = \frac{\omega^{\textrm{att}}_{\textrm{HP}} s}{s + \omega^{\textrm{att}}_{\textrm{HP}}} 
    \cdot
    \frac{\omega^{\textrm{att}}_{\textrm{LP}}}{s + \omega^{\textrm{att}}_{\textrm{LP}}}.
\end{gather}
Given that ${}^{\mathcal{I}}\vec{g} = -g \, {}^{\mathcal{I}}\hat{e}_z$, an estimate of the Euler angles $\hat\theta_{\textrm{acc}}$ and $\hat\phi_{\textrm{acc}}$ can be obtained as
\begin{alignat}{2}
    \label{eq:SteadyStateRollEstimate}
    \hat\phi_{\textrm{acc}} &= \arctan\!2 \Big( & -{}^{\mathcal{B}}{\vec{\hat{g}}}\transpose {}^{\mathcal{B}}{\vec{e}}_y &, \, -{}^{\mathcal{B}}{\vec{\hat{g}}}\transpose {}^{\mathcal{B}}{\vec{e}}_z \Big), \\
    \label{eq:SteadyStatePitchEstimate}
    \hat{\theta}_{\textrm{acc}} &= \arctan\!2 \Big( & {}^{\mathcal{B}}{\vec{\hat{g}}}\transpose {}^{\mathcal{B}}{\vec{e}}_x &, \, \sqrt{ ({}^{\mathcal{B}}{\vec{\hat{g}}}\transpose {}^{\mathcal{B}}\hat{e}_y)^2 + ({}^{\mathcal{B}}{\vec{\hat{g}}}\transpose {}^{\mathcal{B}}\hat{e}_z)^2 } \Big).
\end{alignat}
Finally, under the no-slip condition, an estimate of the yaw angle can be obtained by comparing the difference between the left and right motor encoder counts,
\begin{equation}
    \label{eq:SteadyStateYawEstimateEncoders}
    \hat\psi_{\textrm{enc}} = \frac{2\pi}{12}\frac{(N^{\textrm{enc}}_{\textrm{R}} - N^{\textrm{enc}}_{\textrm{L}})}{i_{\textrm{gb}}}\frac{r_w}{d},
\end{equation}
where $d$ represents the distance between the wheels along the inertial $y$-axis.

The Euler angle rate estimates $[\hat{\dot\phi}_{\textrm{gyr}}, \, \hat{\dot\theta}_{\textrm{gyr}}, \, \hat{\dot\psi}_{\textrm{gyr}}]\transpose$ are obtained by projecting the measured body angular velocities to the relevant (intermediate) axes of the body frame $\{\mathcal{B}\}$,
\begin{equation}
     \begin{bmatrix}
         \hat{\dot\phi}_{\textrm{gyr}} \\
         \hat{\dot\theta}_{\textrm{gyr}} \\
         \hat{\dot\psi}_{\textrm{gyr}}
     \end{bmatrix}
     = 
     \begin{bmatrix}
         1  & \sin(\hat\phi) \tan(\hat\theta)\phantom{^{-1}}  & \cos(\hat\phi) \tan(\hat\theta)\phantom{^{-1}} \\
         0  &  \cos(\hat\phi)\phantom{^{-1}}                  &  -\sin(\hat\phi)\phantom{^{-1}} \\ 
         0  &  \sin(\hat\phi)\big(\cos(\hat\theta)\big)^{-1} & \cos(\hat\phi)\big(\cos(\hat\theta)\big)^{-1} \\
    \end{bmatrix}
    {}^{\mathcal{B}}\vec{\omega}_{\,\nicefrac{\mathcal{B}}{\mathcal{I}}}^{\,\textrm{gyr}}.
    \label{eq:EulerAngleRates_Gyroscope}
\end{equation}

For each of the Euler angles, a total estimate, $\hat\alpha \in ( \hat\phi, \hat\theta, \hat\psi) $, is obtained by fusing the corresponding steady-state estimate, $\hat\alpha_{\textrm{ss}} \in ( \hat\phi_{\textrm{acc}}, \hat\theta_{\textrm{acc}}, \hat\psi_{\textrm{enc}})$, and gyrometer-based, derivative estimate, $\hat{\dot\alpha}_{\textrm{gyr}} \in ( \hat{\dot\phi}_{\textrm{gyr}}, \hat{\dot\theta}_{\textrm{gyr}}, \hat{\dot\psi}_{\textrm{gyr}})$, respectively, with a bias removal system according to
\begin{align}
    \label{eq:ComplimentaryFilter_EstimateFusion}
    \hat\alpha &= \frac{\tau_{\alpha}}{s + \tau_{\alpha}} \hat\alpha_{\textrm{ss}} + \frac{1}{s + \tau_{\alpha}} \big( \hat{\dot\alpha}_{\textrm{gyr}} - b_{\hat{\dot\alpha}_{\textrm{gyr}}}\big), \\
    \label{eq:ComplimentaryFilter_BiasDetermination}
    b_{\hat{\dot\alpha}_{\textrm{gyr}}} &= \frac{k_{\alpha}}{s}\big(\hat\alpha - \hat\alpha_{\textrm{ss}}\big).
\end{align}
Additionally, an estimate of the Euler angle rates is obtained as
\begin{equation}
    \label{eq:ComplimentaryFilter_RateCalculation}
    \hat{\dot\alpha} = \hat{\dot\alpha}_{\textrm{gyr}} - b_{\hat{\dot\alpha}_{\textrm{gyr}}}.
\end{equation}
All filters are discretized using the Tustin transform and implemented using the parameters of \cref{tab:StateEstimatorParameters}. The value of all complimentary filter gains was chosen to be as high as possible without inducing an excessive amount of phase lag.

To correctly determine the value of ${}^{\mathcal{B}}\vec{\omega}_{\,\nicefrac{\mathcal{B}}{{\mathcal{I}}}}$ in \cref{eq:BodyFrameGravityVectorEstimate} without any potential bias present in the measurement of the 3-axis gyroscope, ${}^{\mathcal{B}}\vec{\omega}^{\textrm{gyr}}_{\,\nicefrac{\mathcal{B}}{{\mathcal{I}}}}$, a calibration process is completed prior to every experiment. During this process, the bias determined by \cref{eq:ComplimentaryFilter_BiasDetermination} is transformed to the 3-axes of the gyroscope by using the inverse relation of \cref{eq:EulerAngleRates_Gyroscope}. These values are subtracted from the measurements of the 3-axis gyroscope to approximate the true body angular velocity in \cref{eq:BodyFrameGravityVectorEstimate}, and are held fixed after the calibration process.

Finally, given the use of the SKROSS Reload 10 power bank instead of the standard AA batteries on the Sigi platform, the robot's center of gravity is offset from the body frame. This offset is measured to be approximately \qty[scientific-notation=false,round-precision=2]{-0.078}{\radian} and is subtracted from the final estimate $\hat\theta$ before being passed onto the control algorithm.

\begin{table}[t]
\centering
\begin{tabular}{lccc}
\hline
\textbf{Quantity} & \textbf{Symbol}   & \multicolumn{1}{c}{\textbf{Value}}    & \textbf{Unit} \\ \hline
Sampling time (of all algorithms) & $T_s$ &  \num[scientific-notation=false,round-precision=1]{0.01} & $\si{\second}$ \\
Position vector of the IMU w.r.t P & ${}^{\mathcal{B}}r_{\nicefrac{\textrm{IMU}}{\textrm{P}}}$ & ${\begin{bmatrix} -\num[scientific-notation=false,round-precision=2]{0.012} \\ -\num[scientific-notation=false,round-precision=2]{0.018} \\ \phantom{-}\num[scientific-notation=false,round-precision=2]{0.066} \end{bmatrix}}$ & $\si{\metre}$ \\
\begin{tabular}[c]{@{}l@{}}Cut-off frequencies band-pass \\ \ \ \ attitude filter\end{tabular} & \begin{tabular}[c]{@{}l@{}}$\omega^{\textrm{att}}_{\textrm{HP}}$ \\ $\omega^{\textrm{att}}_{\textrm{LP}}$ \end{tabular} & \begin{tabular}[c]{@{}l@{}} $0.5 \, (\nicefrac{2\pi}{T_s})$ \\ $0.7 \, (\nicefrac{2\pi}{T_s})$ \end{tabular} & \begin{tabular}[c]{@{}l@{}} $\si{\radian\per\second}$ \\ $\si{\radian\per\second}$ \end{tabular} \\
\begin{tabular}[c]{@{}l@{}}Cut-off frequency low-pass \\ \ \ \ translational velocity filter\end{tabular} & $\omega^{\textrm{vel}}_{\textrm{LP}}$ & \num[scientific-notation=false,round-precision=2]{50} & $\si{\radian\per\second}$ \\
Roll complementary filter gains & $(\tau_\phi, k_{\phi})$ & $(\num[scientific-notation=false,round-precision=1]{1}, \, \num[scientific-notation=false,round-precision=1]{0.1})$ & 1\\ 
Pitch complementary filter gains & $(\tau_\theta, k_{\theta})$ & $(\num[scientific-notation=false,round-precision=1]{0.5}, \, \num[scientific-notation=false,round-precision=1]{0.01})$ & 1\\ 
Yaw complementary filter gains & $(\tau_\psi, k_{\psi})$ & $(\num[scientific-notation=false,round-precision=1]{1}, \, \num[scientific-notation=false,round-precision=1]{0.1})$ & 1\\ 
\hline
\end{tabular}
\caption{An overview of the parameters of the state estimator }
\label{tab:StateEstimatorParameters}
\end{table}

\subsection{Position and Velocity Estimation}
\label{sec:StateEstimation_Position}
To determine the robot's translational position and velocity, the encoders mounted on the motor shafts are used. This signal is unwrapped, averaged and multiplied by the total gear ratio to obtain the relative angle between the wheel and pendulum body,
\begin{equation}
    \label{eq:RelativePitchAngleCalculation}
    \Delta \theta_{\textrm{enc}} = \frac{1}{i_{\textrm{gb}}}\frac{2\pi}{12}\bigg(\frac{N^{\textrm{enc}}_{\textrm{R}} + N^{\textrm{enc}}_{\textrm{L}}}{2}\bigg).
\end{equation}
To estimate the translational position $x_w$, the relative angle of the pendulum to the wheel is corrected,
\begin{equation}
    \hat{x}_w = r_w \bigg( \hat \theta + \Delta \theta_{\textrm{enc}}\bigg).
    \label{eq:x_w_EncoderEstimate}
\end{equation}
The translational-velocity estimate $\hat{\dot{x}}_w$ is computed analogously using a first-order difference of the scaled encoder count $\Delta \theta_{\textrm{enc}}$ and the estimated pitch angle rate, $\hat{\dot{\theta}}$. Before correcting for the relative motion of the pendulum body, the first-order difference of the scaled encoder count is passed through a first-order low-pass filter, implemented using the Tustin transform, with a cut-off frequency of $\omega^{\textrm{vel}}_{\textrm{LP}}$ to minimize the effects of the backlash present in the gearbox. 

The complete state estimator is shown as a block diagram in \cref{fig:CompleteControlDiagram}.

\section{Control-Oriented Model}
\label{sec:ControlOrientedModel}
Given the complex, nonlinear system model of the Sigi platform, a control-oriented model is set up by linearizing the complete, deadband-compensated \acrshortpl{EoM} at the (unstable) equilibrium position $x = 0$. The value of certain parameters in the (linearized) \acrshortpl{EoM} are known or easily obtained via measurements, such as
\begin{itemize}
    \item the gearbox ratio, $i_{\textrm{gb}}$,
    \item the radius, distance between, mass and mass moment of inertia of each wheel, $r_w$, $d$, $m_w$, $J_w$, respectively,
    \item and the mass of the pendulum body, $m_p$.
\end{itemize}
The remaining parameters are either not directly measurable, and/or are influenced by the backlash present in the gearbox. As a result, the motor constant $K$ and motor resistance $R$, and the pendulum body mass moment of inertia $J_p$ and \acrshort{CoG} offset $l_c$ are determined experimentally to obtain a control-oriented model. 

\subsection{Motor Parameter Identification}
\label{sec:ControlOrientedModel_MotorParameters}
To determine the motor constant $K$ and motor resistance $R$, firstly, the voltage deadband $u_0$ is determined experimentally. For both motors this is found to be approximately $0.625 \pm 0.1\,\si{\volt}$. 

Next, by flipping the Sigi platform upside down, neglecting the inertia of the gearbox and applying a voltage input on top of the voltage deadband $u_0$, the free-spinning wheel is used as a known load. Accordingly, the motor speed $\omega$ is described by
\begin{equation}
    \label{eq:CTMotorConstantIdentification}
    \frac{J_w}{i_{\textrm{gb}}^2}\dot{\omega} = \frac{K}{R}u - \frac{K^2}{R}\omega.
\end{equation}
To identify the values of $K$ and $R$, an open-loop input voltage signal $(u(k))_{k=1}^N$ is generated by considering a $9$\textsuperscript{th} order, full-length \acrfull{PRBS} signal for $8$ periods at a sampling rate of \qty[scientific-notation=false,round-precision=2]{20}{\hertz}. This signal is scaled such that $\textrm{db}_{u_0}\left(u(k)\right) \in [\qty[scientific-notation=false]{0.7125}{\volt}, \qty[scientific-notation=false]{1.9125}{\volt}]$. To capture the dynamics primarily in a frequency range of interest, the \acrshort{PRBS} signal is re-sampled at \qty[scientific-notation=false,round-precision=2]{100}{\hertz} and passed through two first-order low-pass filters with a cutoff frequency of \si{\frac{70}{2\pi}}\,\unit{\hertz} and one first-order high-pass filter with a cutoff frequency of \si{\frac{30}{2\pi}}\,\unit{\hertz}.

Discretizing \cref{eq:CTMotorConstantIdentification} with an Euler forward discretization and applying the aforementioned open-loop voltage signal, the values of $K$ and $R$ are determined using a least-squares regression. The identified values of $K$ and $R$ are shown in \cref{tab:SigiModelParameters}.

\subsection{Pendulum Body Parameter Identification}
\label{sec:sec:ControlOrientedModel_PendulumBody}
Next, given the identified motor constants, the pendulum body mass moment of inertia $J_p$ and \acrshort{CoG} offset
$l_c$ are determined. This is done by clamping the wheels of the Sigi platform, such that $x_w = \dot{x}_w = \ddot{x}_w = 0$. Neglecting the backlash present in the motor gearbox, applying a deadband compensation and linearizing the resulting \acrshortpl{EoM}, the dynamics are governed by
\begin{equation}
    \label{eq:LinearizedPendulumEoM}
    (J_p + m_pl_c^2) \ddot{\theta} = -\frac{2K}{R}i_{\textrm{gb}}u -\frac{2K^2}{R}i_{\textrm{gb}}^2\dot{\theta} + m_p g l_c \theta.
\end{equation}
Given the unstable nature of this system, a state feedback controller is designed via first-principle estimates of the parameters $J_p$, $l_c$ to stabilize the closed-loop system. Next, based on a full-length, $9$\textsuperscript{th} order \acrshort{PRBS} for $3$ periods at a sampling rate of \qty[scientific-notation=false,round-precision=2]{10}{\hertz}, a reference angle trajectory between $[-2.5 \, (\nicefrac{\pi}{180}), \, \allowdisplaybreaks 2.5 \, (\nicefrac{\pi}{180})]$ \unit{\radian}

is generated. Given the nonlinear nature of the backlash present in the gearbox, a linear model capturing the dynamics primarily in the frequency range of interest is sought by passing the scaled \acrshort{PRBS} signal through two first-order low-pass filters with a cutoff frequency of \si{\frac{15}{2\pi}}\,\unit{\hertz} and one first-order high-pass filter with a cutoff frequency of \si{\frac{5}{2\pi}}\,\unit{\hertz}.

A voltage signal $(u(k))_{ k=1}^{N}$ is generated by the aforementioned state feedback controller by setting up a closed-loop system to track the generated reference angle. Discretizing \cref{eq:LinearizedPendulumEoM} with an Euler forward discretization, a least-squares regression is again used to determine the values of the parameters $J_p$ and $l_c$. The identified values of $J_p$ and $l_c$ are shown in \cref{tab:SigiModelParameters}.

\subsection{Linearization}
\label{sec:ControlOrientedModel_Linearization}
Using the parameter values of \cref{tab:SigiModelParameters} and assuming the deadband voltage $u_0$ is compensated using an additional feedforward input, the \acrshortpl{EoM} are linearized and discretized using a zero-order hold, which leads to the discrete-time \acrfull{LTI} model $x^+ = f(x, u)$
\begin{equation}
    \label{eq:SigiDTLTImodel}
    \begin{aligned}
        = 
        \underbrace{\begingroup
            \sisetup{print-exponent-implicit-plus=true}
            \sisetup{tight-spacing=true}
            \sisetup{retain-zero-exponent=false}
            \setlength{\arraycolsep}{2pt} 
            \begin{bNiceMatrix}
                \num[scientific-notation=false,round-precision=1]{1.00e+00} \,\, & \num{9.8779e-03}           & -\num{3.8056e-05}                   & \num{4.7540e-06}       \\
                \num[scientific-notation=false,round-precision=1]{0} \,\,        & \num{9.7583e-01}           & -\num{7.4328e-03}                   & \num{9.2858e-04}       \\
                \num[scientific-notation=false,round-precision=1]{0} \,\,        & \num{4.6068e-03}           & \phantom{-}\num{1.0047e+00}e\!+\!0  & \num{9.8313e-03}       \\
                \num[scientific-notation=false,round-precision=1]{0} \,\,        & \num{9.1259e-01}           & \phantom{-}\num{9.2604e-01}         & \num{9.6816e-01}       \\
            \end{bNiceMatrix}
        \endgroup}_{A}
        x + 
        \underbrace{\begingroup
            \sisetup{print-exponent-implicit-plus=true}
            \sisetup{tight-spacing=true}
            \sisetup{retain-zero-exponent=false}
            \begin{bNiceMatrix}
                \phantom{-}\num{6.7532e-05}      \\
                \phantom{-}\num{1.3370e-02}      \\
                -\num{2.5488e-03}      \\
                -\num{5.0491e-01}      \\
            \end{bNiceMatrix}
        \endgroup}_{B}
        u.
    \end{aligned}
\end{equation}

\section{Controller Synthesis}
\label{sec:ControllerSynthesis}
Using the discrete-time \acrshort{LTI} model of the Sigi platform, three controllers are synthesized: a baseline \acrshort{LQR}, a robust, tube-based \acrshort{MPC} and an \acrshort{NNC}. Of these, only the \acrshort{LQR} and \acrshort{NNC} are capable of running on the Sigi platform's microprocessor in real time and will be used to record empirical control performance data.

\subsection{Linear-Quadratic Regulator}
\label{sec:ControllerSynthesis_LQR}
Given the state estimator and discrete-time \acrshort{LTI} model of the Sigi platform as outlined in \cref{sec:StateEstimation,sec:ControlOrientedModel}, respectively, an \acrshort{LQR} is designed as a baseline controller to stabilize the system. The state and control cost matrices $Q$ and $R$, respectively, of the cost function
\begin{equation}
    \label{eq:DT_LQR_InfiniteHorizonCost}
    \sum_{k=0}^{\infty} x(k)\transpose Q x(k) + u(k)\transpose R u(k)
\end{equation}
associated with the infinite-horizon \acrshort{LQR} synthesis problem are chosen to represent
\begin{align}
    \label{eq:StateCostMatrixDefinition}
    x(k)\transpose Q x(k) &= 25 x_{\textrm{CoG}}(k)^2 + 25 \dot{x}_{\textrm{CoG}}(k)^2, \\
    \label{eq:InputCostMatrixDefinition}
    u(k)\transpose R u(k) &= 7.5 u(k)^2,
\end{align}
where
\begin{align}
    x_{\textrm{CoG}}(k) &= x_{w}(k) + \frac{m_p}{m_p + 2m_w}l_c \theta(k), \\
    \dot{x}_{\textrm{CoG}}(k) &= \dot{x}_{w}(k) + \frac{m_p}{m_p + 2m_w}l_c \dot{\theta}(k).
\end{align}

Using \cref{eq:SigiDTLTImodel} and defining the resulting optimal closed-loop system as $A + BK_{\textrm{LQR}}$ leads to the control gains
\begin{equation}
    \label{eq:LQR_controller}
    K_{\textrm{LQR}} = \begin{bmatrix} \num[round-precision=3]{1.6741e+00} & \num[round-precision=3]{5.1671e+00} & \num[round-precision=3]{4.7773e+00} & \num[scientific-notation=false,round-precision=3]{0.45641} \end{bmatrix}.
\end{equation}
The voltage constraints to the electric motors are guaranteed by saturating the controller output at $\pm\qty[scientific-notation=false,round-precision=2]{2.0}{\volt}$.

\subsection{Robust, Tube-Based Model Predictive Controller}
\label{sec:ControllerSynthesis_RobustTubeMPC}
Next, the procedure to obtain a robust, tube-based \acrshort{MPC} is described. As this controller will be used to generate input-output pairs $\big(x(k),u_{\textrm{MPC}}(x(k))\big)$ for the training of an \acrshort{NNC}, $\varphi$, the (minimal) robustness margin of the \acrshort{MPC} is chosen with this in mind \cite{mybibfile:Dubach2022}. Namely, it follows that after completion of the training process, the \acrshort{NNC} approximately represents a stabilizing \acrshort{MPC} for the dynamic model of \cref{eq:SigiDTLTImodel}. By viewing the deviations of the neural network from this \acrshort{MPC} as a bounded disturbance $w(k) \in \mathcal{W}$ it follows
\begin{equation}
    \varphi(x(k)) = u_{\textrm{MPC}}(x(k)) + w(k), \ \ w(k) \in \mathcal{W}.
\end{equation}
By assuming an a priori bound on the size of the disturbance set $\mathcal{W}$, the open-loop system for which the robust \acrshort{MPC} should be designed,
\begin{equation}
    \label{eq:ClosedLoop_withNoise}
    x(k+1) = Ax(k) + Bu(k) + Bw(k), \ \ w(k) \in \mathcal{W},
\end{equation}
is defined.
Given this bound, a robust, tube-based \acrshort{MPC} is then defined, such that $u_{\text{MPC}}\left(x(k)\right) = K_{\text{tube}}\left(x(k) - z_1\right) + v_1$ with $z_1$ and $v_1$ the solutions to
\begin{subequations}
    \label{eq:RobustMPC_OptimizationProblem}
	\begin{alignat}{6}
		&& \underset{ \substack{\{z_i\}_{i = \{1, \dotsc, N_{\textrm{pred}}+1\}} \\ \{v_i\}_{i = \{1, \dotsc, N_{\textrm{pred}}\} }}}{\textrm{minimize:}} \ && 	\sum_{i=1}^{N_{\textrm{pred}}} z_i\transpose Q z_i + v_i\transpose R v_i + z_{N_{\textrm{pred}}+1}\transpose P z_{N_{\textrm{pred}}+1} \span \span \span \span \span \nonumber \\
        \label{eq:RobustMPC_NominalDynamics}
		&& \text{s.t.} \span \ & z_{i+1} &= \ && Az_{i}+Bv_{i}, \quad && \forall i\in [N_{\textrm{pred}}], 
		\\
        \label{eq:RobustMPC_TightenedStateConstraints}
        && 				   && z_{i} &\in \ && \mathcal{X} \ominus \mathcal{E}, \quad && \forall i\in [N_{\textrm{pred}}], 
        \\
        \label{eq:RobustMPC_TightenedInputConstraints}
		&& 				   && v_{i} &\in \ && \mathcal{U} \ominus K_{\text{tube}} \mathcal{E}, \quad && \forall i\in [N_{\textrm{pred}}], 
		\\
		&& 				   && x(k) - z_{1} &\in \ && \mathcal{E}, \quad &&	\\
		&& 				   && z_{N+1} &\in \ && \mathcal{X}_f. \quad &&
	\end{alignat}
\end{subequations}
Here $Q$ and $R$ are defined as in \cref{eq:StateCostMatrixDefinition,eq:InputCostMatrixDefinition}, respectively, and $P$ is equal to the solution of the discrete-time algebraic Ricatti equation used to solve the infinite-horizon \acrshort{LQR} problem of \cref{sec:ControllerSynthesis_LQR}. In addition, polytopic sets $\mathcal{X}$, $\mathcal{U}$ define the set of allowable states and inputs, respectively, and $\mathcal{X}_f$ defines the maximal invariant set under the \acrshort{LQR} of \cref{eq:LQR_controller} and the tightened state and input constraints of \cref{eq:RobustMPC_TightenedStateConstraints,eq:RobustMPC_TightenedInputConstraints}. With $\mathcal{E} = \allowdisplaybreaks \{ e \in \realsN{4} \mid \sqrt{e\transpose P_{\textrm{tube}} e} \leq \delta_{\textrm{tube}} \}$ representing an ellipsoidal, robust positive invariant set approximating the minimal robust positive invariant set $\oplus_{i=0}^{\infty} (A+BK_{\text{tube}})^iB\mathcal{W}$, the control input $u_\textrm{MPC}(x(k))$ as defined by optimization problem \cref{eq:RobustMPC_OptimizationProblem} is guaranteed to robustly stabilize the system of \cref{eq:ClosedLoop_withNoise} to the set $\mathcal{E}$ \cite{mybibfile:Rawlings2017}.

The feedback controller $K_{\textrm{tube}}$, with minimum exponential decay rate $\rho$, and corresponding robust positive invariant set $\mathcal{E}$ are defined by
\begin{align}
    \label{eq:TubeControllerGains}
    K_{\textrm{tube}} &= \begingroup \begin{bNiceMatrix} \num[round-precision=3]{1.3491e+04} & \num[round-precision=3]{1.3152e+03} & \num[round-precision=3]{4.4497e+02} & \num[round-precision=3]{3.6247e+01} \end{bNiceMatrix} \endgroup, \\
    \label{eq:TubeQuadraticMatrixValue}
    P_{\textrm{tube}} &= \begingroup
        \sisetup{print-exponent-implicit-plus=true}
        \sisetup{tight-spacing=true}
        \sisetup{retain-zero-exponent=false}
        \begin{bNiceMatrix}
            \num{3.6194e+09}                & \num{2.4436e+08}           & \num{1.0542e+08}             & \num{6.5513e+06}       \\
            *                               & \num{1.8630e+07}           & \num{7.2733e+06}             & \num{5.0052e+05}       \\
            *                               & *                          & \num{3.0857e+06}             & \num{1.9514e+05}       \\
            *                               & *                          & *                            & \num{1.3451e+04}       \\
        \end{bNiceMatrix}
    \endgroup, \\
    \label{eq:TubeRadius}
    \delta_{\textrm{tube}} &= \num[scientific-notation=false]{0.69555}.
\end{align}
 These values were found by solving a \acrfull{SDP} attempting to minimize the resulting constraint tightening of the robust, tube-based \acrshort{MPC} of \cref{eq:RobustMPC_OptimizationProblem} \cite{mybibfile:Kolmanovsky1998,mybibfile:Boyd1994}.

Together with \cref{eq:SigiDTLTImodel} and \cref{tab:TubeMPCParameters}, this completely defines the robust, tube-based \acrshort{MPC} of optimization problem \cref{eq:RobustMPC_OptimizationProblem}.

\begin{table}[t]
\centering
\begin{tabular}{lccc}
\hline
\textbf{Quantity} & \textbf{Symbol}   & \multicolumn{1}{c}{\textbf{Value}\phantom{e3}}    & \textbf{Unit} \\ \hline
Max. abs. translational position & $|x_w|_{\textrm{max}}$ & \phantom{0}\num[scientific-notation=false,round-precision=2]{0.10}\phantom{0e3} & $\si{\metre}$ \\
Max. abs. translational velocity & $|\dot{x}_w|_{\textrm{max}}$ & \phantom{0}\num[scientific-notation=false,round-precision=2]{0.25}\phantom{0e3} & $\si{\metre\per\second}$ \\
Max. abs. pitch angle & $|\theta|_{\textrm{max}}$ & \phantom{0}\num[scientific-notation=false,round-precision=3]{0.2182}\phantom{e3} & $\si{\radian}$ \\
Max. abs. pitch rate & $|\dot{\theta}|_{\textrm{max}}$ & \num[scientific-notation=false,round-precision=2]{6.0}\phantom{0e3}  & $\si{\radian\per\second}$ \\
Max. abs. \acrshort{CoG} position & $|x_{\textrm{CoG}}|_{\textrm{max}}$ & \phantom{0}\num[scientific-notation=false,round-precision=2]{0.075}\phantom{e3} & $\si{\metre}$ \\
Max. abs. \acrshort{CoG} velocity & $|\dot{x}_{\textrm{CoG}}|_{\textrm{max}}$ & \phantom{0}\num[scientific-notation=false,round-precision=3]{0.225}\phantom{e3}& $\si{\metre\per\second}$ \\
Max. abs. input voltage & $|u|_{\textrm{max}}$ & \phantom{0}\num[scientific-notation=false,round-precision=2]{2.0}\phantom{00e3}  & $\si{\volt}$ \\
Max. abs. disturbance input & $|w|_{\textrm{max}}$ & \phantom{0}\num[scientific-notation=false,round-precision=2]{0.075}\phantom{e3} & $\si{\volt}$ \\
Min. exponential stability of $K_{\textrm{tube}}$ & $\rho$ & \phantom{0}\num[scientific-notation=false,round-precision=3]{0.815}\phantom{e3} & $1$ \\
Penalty term for slack variables $\varepsilon_i$ & $\rho_\varepsilon$ & \phantom{0}\num[scientific-notation=true,round-precision=3]{5e3}\phantom{0} & $1$ \\
Prediction horizon & $N_{\textrm{pred}}$ & \num[scientific-notation=false,round-precision=2]{30}\phantom{e3} & $1$ \\
\hline
\end{tabular}
\caption{An overview of the parameters defining \cref{eq:RobustMPC_OptimizationProblem}.}
\label{tab:TubeMPCParameters}
\end{table}

\subsection{Neural-Network-Based Controller}
\label{sec:ControllerSynthesis_NNC}
The robust, tube-based \acrshort{MPC} of \cref{sec:ControllerSynthesis_RobustTubeMPC} is not capable of running in real time on the Sigi platform's embedded hardware. Therefore, this section presents a generalizable procedure to synthesize an \acrshort{NNC} imitating the robust, tube-based \acrshort{MPC} of \cref{sec:ControllerSynthesis_RobustTubeMPC} by means of supervised learning. This procedure utilizes knowledge of the underlying \acrshort{MPC} formulation and consists of dataset generation, training and postprocessing steps.

\subsubsection{Dataset generation}
\label{sec:ControllerSynthesis_NNC_DatasetGeneration}
To obtain an \acrshort{NNC} approximating the solutions of \acrshort{MPC} problem \cref{eq:RobustMPC_OptimizationProblem}, training data $\mathcal{D} = \{ (\bar{x}_i, \, u(\bar{x}_i)) \}_{i=1}^N$
is generated. The training data $\mathcal{D}$ is generated by solving optimization problem \cref{eq:RobustMPC_OptimizationProblem} with softened constraints, which has the heuristic justification of allowing a greater amount of training data to capture the strong nonlinearities at the edges of the \acrshort{MPC} problem's feasible set. This constraint softening is achieved by introducing a scaled slack variable $\varepsilon_i$ on the right-hand side of all tightened state, input and terminal set constraints of timestep $i$, for all $i \in [N_{\textrm{pred}}]$. The factors scaling $\varepsilon_i$ are chosen such that a violation of any of the halfspace constraints by $\Delta z_\varepsilon = [\qty[scientific-notation=false,round-precision=1]{0.01}{\metre},\, \qty[scientific-notation=false,round-precision=2]{0.25}{\metre\per\second},\, \nicefrac{0.5\pi}{180}\,\unit{\radian},\, \qty[scientific-notation=false,round-precision=1]{1}{\radian\per\second}]$ or $\Delta v_\varepsilon = \qty[round-precision=2]{3.33e-04}{\volt}$ increases value of $\varepsilon_i$ by $1$. Additionally, a penalty term of $\sum_{i=1}^N \rho_\varepsilon \varepsilon_i^2$ is added to the objective function of the optimization problem. 

The inputs $\bar{x}_i$ of the training dataset are selected from a scaled, four-dimensional hypercube centered at the origin, $\{ \bar{x} \in \realsN{4} \mid \|\bar{x}\|_\infty \leq 1 \}$. Using the maximum allowable absolute values given in \cref{tab:TubeMPCParameters}, the relation between inputs $\bar{x}$ and states $x$ of \cref{eq:SigiDTLTImodel} is defined as $x = D_x \bar{x}$, with $D_x = \textrm{diag}\big(\nicefrac{1}{10}, \, \nicefrac{1}{4}, \, \nicefrac{12.5 \pi}{180}, \, 6\big)$. 

From the formulation of the (softened) optimization problem \cref{eq:RobustMPC_OptimizationProblem} and equations \cref{eq:LQR_controller,eq:TubeControllerGains}, it is known a priori that a large nonlinearity in the \acrshort{MPC} is present at the origin of the state space. Therefore, along each dimension, approximately $30\%$ of the training dataset's inputs correspond to states that lie inside robust positive invariant set $\mathcal{E}$, with the remainder being spaced linearly up to the maximum absolute value allowed by the values given in \cref{tab:TubeMPCParameters}. The training data used in this work consists of \num[scientific-notation=false,round-precision=7]{2288807} samples of the \acrshort{MPC} solution.

\subsubsection{Neural network training}
\label{sec:ControllerSynthesis_NNC_NNTraining}
The \acrshort{NNC} is restricted to be an $\ell$-layer feedforward neural network $\varphi\colon \realsN{n} \mapsto \realsN{n_u}$ described in full generality by
\begin{subequations}
    \label{eq:OverallFFNeuralNet}
    \begin{alignat}{2}
        \label{eq:FFCompostionNeuralNet}
        \varphi(x) &= f^{\varphi}_{\ell+1} \circ \phi_\ell \circ f^{\varphi}_{\ell} \circ \, \ldots \circ \phi_1 \circ f^{\varphi}_1(x), \ && \\
        f^{\varphi}_i(x) &= W_i x + b_i. \quad && 
        \label{eq:MatrixMultiplicationNeuralNet}
    \end{alignat}
\end{subequations}
Here, $W_i \in \realsN{n_{i} \times n_{i-1}}$, $b_i \in \realsN{n_i}$, $\phi_i \colon \realsN{n_i} \mapsto \realsN{n_{i}}$ represent the weights, biases and stacked activation functions of layer $i$, respectively, and $n_i$ denotes the number of inputs to (hidden) layer $i+1$. The \acrshort{NNC} examined in this work uses solely \acrshort{ReLU} activation functions and has layer widths of $(n_0, \, n_1, \, n_2, \, n_3, \, n_4) = \allowdisplaybreaks (4, \, 18, \, 12, \, 6, \, 1).$ This type of architecture has (empirically) been shown to generate neural networks with a large number of piecewise-affine regions \cite{mybibfile:Serra2018}. 

A batch-normalized \acrfull{MSE} loss function and the Adam optimizer \cite{mybibfile:Kingma2017} are used to train the neural network. An adaptive learning rate is used during the training process, whereby the learning rate is initialized at \num[scientific-notation=false,round-precision=1]{0.01} and decreases to $[ \num[scientific-notation=false,round-precision=1]{0.005} \ \num[scientific-notation=true,round-precision=1]{0.0005} \ \num[scientific-notation=true,round-precision=1]{0.00001}]$ after the $50$\textsuperscript{th}, $125$\textsuperscript{th}, $225$\textsuperscript{th} epoch. The overall training consists of \num[scientific-notation=false,round-precision=3]{425} epochs, with a batch size of \num[scientific-notation=false,round-precision=3]{1000}. 

\subsubsection{Postprocessing}
\label{sec:ControllerSynthesis_NNC_Postprocessing}
Given the \acrshort{NNC} trained using the above procedure, it must be ensured that the origin is an equilibrium point of the closed-loop system, i.e. $\varphi(0) = 0$. This is achieved by (minimally) adapting the weights and biases of the output layer, $W_{\ell+1}$, $b_{\ell+1}$, to $W^{\textrm{new}}_{\ell+1}$, $b^{\textrm{new}}_{\ell+1}$, respectively. Let $\lambda^0_\ell$ represent the output of the final hidden layer for a network input of zero, i.e. $\lambda^0_\ell = \phi_\ell \circ f^{\varphi}_{\ell} \circ \, \ldots \circ \phi_1 \circ f^{\varphi}_1(0)$. Then, new weights and biases for the output layer are found by solving
\begin{subequations}
	\label{eq:NNRemoveOffsetOptimizationProblem}
	\begin{alignat}{3}
		&& \underset{ \substack{W^{\textrm{new}}_{\ell+1}, \, b^{\textrm{new}}_{\ell+1} } }{\textrm{minimize:}} \ && 	\left\| \begin{matrix}
		    (W^{\textrm{new}}_{\ell+1} - W_{\ell+1}) \oslash {W_{\ell+1}} & b^{\textrm{new}}_{\ell+1} - b_{\ell+1}
		\end{matrix}\right\|_{\infty} \span \nonumber \\
		&& \text{s.t.} \span \ & {W^{\textrm{new}}_{\ell+1}} \lambda^0_{\ell} + b^{\textrm{new}}_{\ell+1} = 0.  \span
	\end{alignat}
\end{subequations}
To conclude the generalizable synthesis procedure, the \acrshort{NNC} is adapted to satisfy the Sigi platform's input voltage constraints of $\pm\qty[scientific-notation=false,round-precision=2]{2.0}{\volt}$ via the addition of two \acrshort{ReLU} neurons, resulting in a final network architecture of $(n_0, \, n_1, \, n_2, \, n_3, \, n_4, n_5, \, n_6) = \allowdisplaybreaks (4, \, 18, \, 12, \, 6, \, 1, \, 1, \, 1)$.

\subsection{Controller Output Modification}
\label{sec:ControllerSynthesis_OutputModification}
As part of the experimental validation in this work, a feedforward input signal $(u_{\textrm{ff}}(k))_{k=1}^N$ corresponding to a regulation or reference-tracking task is required. In addition, the control-oriented model of \cref{eq:SigiDTLTImodel} is derived under the assumptions that
\begin{itemize}
    \item the deadband voltage $u_0$ is compensated, 
    \item the Sigi platform performs only planar movements and thus has a constant yaw angle.
\end{itemize} 
To guarantee the validity of these assumptions and generate a valid feedforward signal for the reference-tracking task, the control outputs for the left and right motors, which are generated by the controllers synthesized in this section,
are modified before being sent to each of the electric motors. 

\subsubsection{Feedforward signal}
A feedforward input signal $(u_{\textrm{ff}}(k))_{k=1}^N$ corresponding to a regulation or reference-tracking task is required. Clearly, for a regulation task, a reference of $r(k) = [0, \, 0, \, 0, \, 0]\transpose$ and feedforward control input $u_{\textrm{ff}}(k) = 0$ for all $k$ are supplied. 

For the reference-tracking task, a feedforward signal \smash{$(u_{\textrm{ff}}(k))^{N}_{k=1}$} for \cref{eq:SigiDTLTImodel} is generated that defines a maneuver consisting of i) a regulation task, ii) a translational movement, and iii) a regulation task at a different translational position. This maneuver is parameterized by

\begin{equation}
    \label{eq:ManeuverReferenceTrajectory}
    r_{\textrm{opt}}(k) = 
    \begin{cases}
        \begin{bmatrix} 0 \phantom{v_x(t_2-t_1)}\!\!\! & 0 \phantom{v_x}\!\! & 0 & 0 \end{bmatrix}\transpose                           & \text{if}\; \nicefrac{t_0}{T_s} \leq k < \nicefrac{t_1}{T_s}, \\
        \begin{bmatrix} v_x(kT_s - t_1) \!\!\!\! & v_x \phantom{0}\!\! & 0 & 0 \end{bmatrix}\transpose                 & \text{if}\; \nicefrac{t_1}{T_s} \leq k < \nicefrac{t_2}{T_s}, \\
        \begin{bmatrix} v_x(t_2-t_1) \phantom{0}\!\!\! & 0 \phantom{v_x}\!\! & 0 & 0 \end{bmatrix}\transpose                           & \text{if}\; \nicefrac{t_2}{T_s} \leq k \leq \nicefrac{t_3}{T_s}.
    \end{cases}
\end{equation}
Defining the error of the state with respect to this reference as $e_{r_{\textrm{opt}}}(k) = r_{\textrm{opt}}(k) - x(k)$, $u_{\textrm{ff}}(k)$ is selected to be of the form
\begin{equation}
    \label{eq:FeedforwardControlInput}
    u_{\textrm{ff}}(k) = 
    \begin{cases}
        0                                                           & \text{if}\; \nicefrac{t_0}{T_s} \leq k < \nicefrac{t_1}{T_s}, \\
        u_{r_{\textrm{opt}}}(k) - K_\textrm{ff}e_{r_{\textrm{opt}}}(k) + u_{\textrm{tran}}(k)       & \text{if}\; \nicefrac{t_1}{T_s} \leq k < \nicefrac{t_2}{T_s}, \\
        0                                                           & \text{if}\; \nicefrac{t_2}{T_s} \leq k \leq \nicefrac{t_3}{T_s}.
    \end{cases}
\end{equation}
Under the assumption that $x(\nicefrac{t_0}{T_s}) = [0, \, 0, \, 0, \, 0]\transpose$, the reference is tracked perfectly under $u_{\textrm{ff}}(k) = 0$ for $\nicefrac{t_0}{T_s} \leq k < \nicefrac{t_1}{T_s}$. Next, for all $\nicefrac{t_1}{T_s} \leq k < \nicefrac{t_2}{T_s}$, the error dynamics under the feedforward control input become
\begin{alignat}{1}
    \label{eq:FeedforwardErrorDynamics}
        e_{r_{\textrm{opt}}}(k+1) &= r_{\textrm{opt}}(k+1) - Ax(k) - Bu_{\textrm{ff}}(k) \\
            & 
            \begin{aligned} 
            \label{eq:SimplifiedFeedforwardErrorDynamics}
                &= \big(A + BK_{\textrm{ff}}\big)e_{r_{\textrm{opt}}}(k) -B u_{\textrm{tran}}(k) \\ 
                & \qquad \quad + r_{\textrm{opt}}(k+1) - Ar_{\textrm{opt}}(k) - Bu_{r_{\textrm{opt}}}(k),
            \end{aligned}
\end{alignat}
with $e_{r_{\textrm{opt}}}(\nicefrac{t_1}{T_s})$ defined by $x(\nicefrac{t_1}{T_s})=[0, \, 0, \, 0, \, 0]\transpose$ given that $u_{\textrm{ff}}(\nicefrac{t_1}{T_s-1})=0$. The term $u_{r_{\textrm{opt}}}(k)$ is constructed to satisfy $r_{\textrm{opt}}(k+1) = Ar_{\textrm{opt}}(k) + Bu_{r_{\textrm{opt}}}(k)$, effectively removing this term from \cref{eq:SimplifiedFeedforwardErrorDynamics}. Next, $K_\textrm{ff} = [\num[scientific-notation=false,round-precision=3]{0.91287}, \, \num[scientific-notation=false,round-precision=3]{4.3439e+00}, \, \num[scientific-notation=false,round-precision=3]{4.6349e+00}, \, \num[scientific-notation=false,round-precision=3]{0.43292} ] $ is selected to be a stabilizing controller for \cref{eq:SigiDTLTImodel}, thereby asymptotically bringing the state to the desired reference. Finally, the term $u_{\textrm{tran}}(k)$ is defined as the minimum energy control input such that $x(t_2/T_s) = [v_x(t_2 - t_1), \, 0, \, 0, \, 0]\transpose$\!\!\!,\hspace{0.3ex} to allow the final piecewise constant reference to be tracked under $u_{\textrm{ff}}(k) = 0$ for $t_2/T_s \leq k \leq t_3/T_s$. Therefore, by definition, $u_\textrm{tran}(k)$ is defined by the solution to the minimum energy control problem,
\begin{subequations}
	\begin{alignat}{6}
		&& \underset{ u_\textrm{tran}(k) }{\textrm{minimize:}} \ && 	\sum_{k = \nicefrac{t_1}{T_s}}^{\nicefrac{t_2}{T_s} - 1} u_\textrm{tran}(k)\transpose u_\textrm{tran}(k) \span \span \span \span \span \nonumber \\
        && \text{s.t.} \span \ & \begin{aligned} & e_\textrm{tran}(k+1) \\ & \phantom{} \end{aligned} & \begin{aligned} &= \\ & \phantom{}  \end{aligned} \ && \begin{aligned} & \big(A + BK_{\textrm{ff}}\big)e_\textrm{tran}(k) -B u_{\textrm{tran}}(k), \\ 
        & \qquad \forall k\in \{ \nicefrac{t_1}{T_s}, \dots, \nicefrac{t_2}{T_s} - 1\},\end{aligned} \quad && 
		\\
        && 				   && e_\textrm{tran}(\nicefrac{t_1}{T_s}) &= \ &&  [0, \, v_x, \, 0, \, 0]\transpose, \quad &&
        \\
		&& 				   && e_\textrm{tran}(\nicefrac{t_2}{T_s}) &= \ &&  [0, \, v_x, \, 0, \, 0]\transpose, \quad &&
	\end{alignat}
\end{subequations}
for which a closed-form solution exists \cite{mybibfile:LaSalle1986}. In this manner, the feedforward signal \smash{$(u_{\textrm{ff}}(k))^{\scriptscriptstyle t_3/T_s}_{\scriptscriptstyle k=t_0/T_s}$}
of \cref{eq:FeedforwardControlInput} defines the state trajectory \smash{$(x(k))^{\scriptscriptstyle t_3/T_s}_{\scriptscriptstyle k=t_0/T_s}$} approximating the maneuver of \cref{eq:ManeuverReferenceTrajectory}, thereby allowing this state trajectory to be used as the reference for all states in the reference-tracking task of \cref{sec:Results}. 

\begin{figure*}[t]
    \centering
    \def\svgwidth{0.965\linewidth}
    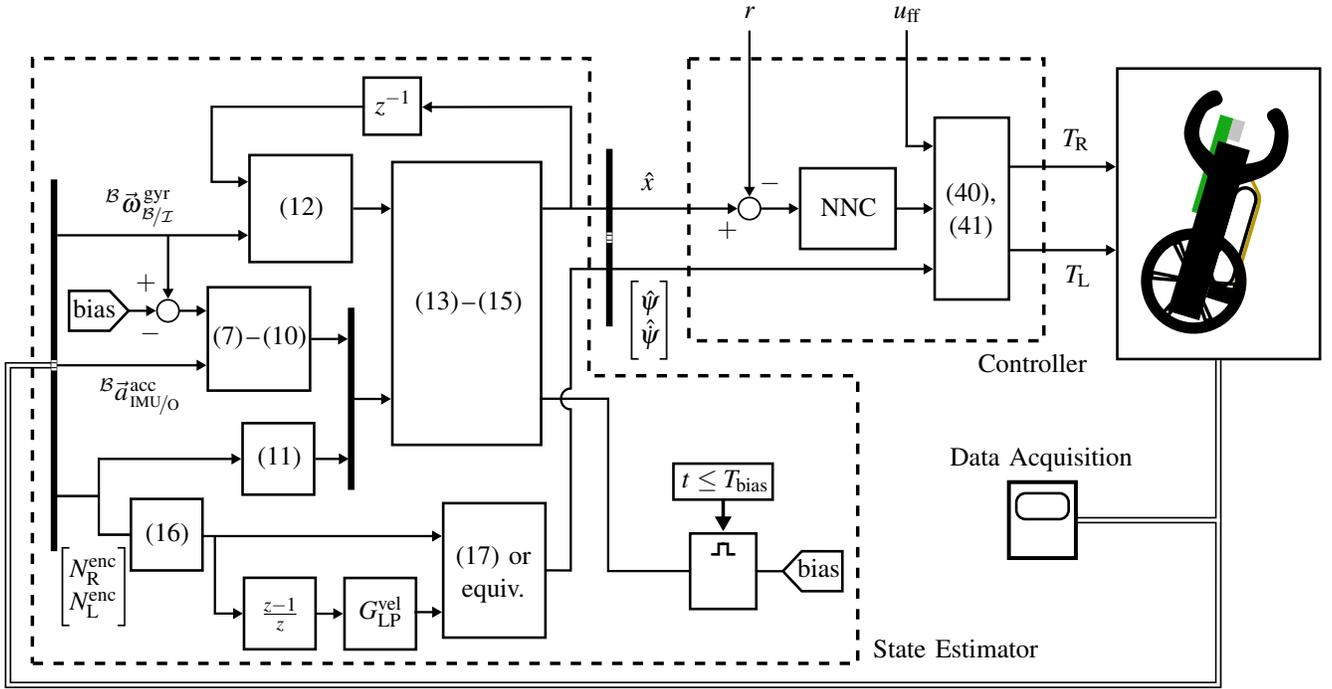
    \caption{A block diagram of the complete control loop used to control the Sigi platform, depicted on the right. The state estimator of \cref{sec:StateEstimation} and complete controller developed in \cref{sec:ControllerSynthesis} are shown.}
    \centering
    \label{fig:CompleteControlDiagram}
\end{figure*}

\subsubsection{Active yaw control}
To guarantee the Sigi platform only performs planar movements, an active yaw controller of the form $\pm K_{\textrm{yaw}} [\hat\psi(k), \, \hat{\dot{\psi}}(k)]\transpose$ is constructed to modulate the voltage setpoints to the left and right electric motors, respectively. The control gain $K_{\textrm{yaw}} = [\num[scientific-notation=false,round-precision=2]{0.78} \ \num[scientific-notation=false,round-precision=2]{0.49} ]$ is determined by examining a linearized model considering the two additional yaw-related \acrshortpl{DoF} \cite{mybibfile:Kim2014}. The required (estimations of) the mass moments of inertia are obtained using approximate dimensions of the wheels and pendulum body and the assumption that the respective masses $2m_w$, $m_p$ are distributed homogeneously throughout this volume.

\subsubsection{Deadband voltage compensation}
Summing the controller output modifications, the total desired voltage becomes
\begin{equation}
    \label{eq:DesiredControlInputCalculation}
    u_{\textrm{des}}(k) = u(k) + u_{\textrm{ff}}(k) \pm K_{\textrm{yaw}} \begin{bmatrix} \hat\psi(k) \\ \hat{\dot{\psi}}(k) \end{bmatrix},
\end{equation}
where $u(k)$ is the control output from any controller of \cref{sec:ControllerSynthesis} evaluated for the (negative) tracking error $\hat{x}(k) - r(k)$. Next, the deadband voltage $u_0$ is compensated by means of an additional input of $\pm u_0$, leading to the total voltage 
\begin{equation}
    \label{eq:DesiredControlInputHysteresisCompensation}
    u_{\textrm{tot}}(k) = 
    \begin{cases}
        0                                                           & \text{if}\; u_{\textrm{des}}(k) = 0, \\
        u_{\textrm{des}}(k) + \textrm{sign}(u_{\textrm{des}}(k))u_0      & \text{otherwise},
    \end{cases}
\end{equation}
being sent to the electric motors of the Sigi platform.

The complete control loop including the control output modifications presented in this section are shown in \cref{fig:CompleteControlDiagram}. These control output modifications allow the assumptions of the control-oriented model of \cref{eq:SigiDTLTImodel} to be met and thereby allow this model to be used in the analysis of closed-loop stability properties of the Sigi platform under the \acrshort{NNC}.
        
\section{Stability Verification}
\label{sec:StabilityVerification}
The stability properties of the closed-loop system formed by using the \acrshort{NNC} of \cref{sec:ControllerSynthesis} to control the open-loop system \cref{eq:SigiDTLTImodel} are unknown. Additionally, as previously noted, given the limited control authority of the actuators on the Sigi platform, there exists no globally-stabilizing controller. Therefore, a certificate of local asymptotic stability and an inner estimate of the \acrshort{ROA} are sought.

\begin{defn}[Local Asymptotic Stability]
    Closed-loop system $x^+ = f\big(x, u(x)\big)$ is \acrfull{LAS} if
    \begin{itemize}
        \item it is stable in the sense of Lyapunov, i.e. for every $\epsilon > 0$, $\exists \delta(\epsilon) > 0$ such that $\|x(0)\|^2 < \delta \implies \|x(k)\|^2 < \epsilon$ for all $k \in \integersN{}_{\geq 0}$.
        \item it is locally attractive, i.e. $\exists \eta > 0$ such that $\lim_{k \to \infty} \|x(k)\| = 0$ for all $\|x(0)\| < \eta$.
    \end{itemize}
\end{defn}

\begin{defn}[Region of Attraction]
    The \acrfull{ROA} of the equilibrium point $x = 0$ of closed-loop system $x^+ = f\big(x, u(x)\big)$ is the set $\mathcal{X}_{\textrm{RoA}}$ of all points such that $x \in \mathcal{X}_{\textrm{RoA}}$ implies $\lim_{k\to\infty} \|x(k)\| = 0$.
\end{defn}

A locally valid Lyapunov function serves as a certificate of \acrshort{LAS}.
\begin{defn}[Lyapunov Function, Def. B.12 \cite{mybibfile:Rawlings2017}]
    \label{def:LocalLyapunovFunction}
    Suppose that $\mathcal{X}$ is positive invariant and the origin $0\subset \mathcal{X}$ is an equilibrium point for $x^+ = f(x,u(x))$. A function $V: \mathcal{X}\rightarrow R_{\geq 0}$ is said to be a Lyapunov function in $\mathcal{X}$ for the system $x^+ = f(x,u(x))$ if there exist functions $\alpha_1, \alpha_2 \in \mathcal{K}_{\infty}$, and continuous, positive definite function $\alpha_3$ such that for any $x \in \mathcal{X}$ 
\end{defn}
\begin{subequations}\label{eq:V_conditions}
    \begin{alignat}{1}
    V(x) &\geq \alpha_1(\|x\|) \label{eq:V_conditions_a}, \\
    V(x) &\leq \alpha_2(\|x\|) \label{eq:V_conditions_b}, \\
    V(x) - V(f(x,u(x))) &\geq \alpha_3(\|x\|)\label{eq:V_conditions_c}.
\end{alignat}
\end{subequations}

Furthermore, the positive invariant set $\mathcal{X}$ of \cref{def:LocalLyapunovFunction} serves as an inner estimate of the closed-loop system's \acrshort{ROA}.

In this work a locally valid Lyapunov function is found in two steps using an \acrshort{SOS}-based stability verification procedure \cite{mybibfile:Korda2022,mybibfile:Newton2022}.
    
Firstly, to satisfy \cref{eq:V_conditions_c}, an \acrshort{SOS} optimization problem is formulated to return a non-negative function $V$ 
satisfying,
\begin{equation}
    \label{eq:DecreaseConditionInformationTerm}
    V(x) - V(x^+) - \|x\|_P^2 - \Gamma\big(x, \, x^+ ; (f, \varphi \circ D_x^{-1}, \mathcal{Q})\big) \geq 0.
\end{equation}
Here $P$ is a positive definite matrix, $\Gamma\big(x, \, x^+ ; (f, \varphi \circ D_x^{-1}, \mathcal{Q})\big)$ is a polynomial term parameterized by $(f, \varphi \circ D_x^{-1}, \mathcal{Q})$ that is guaranteed to be non-negative for all states satisfying i) $x^+ = f\big(x, \varphi \circ D_x^{-1} (x)\big)$ and ii) $x \in \mathcal{Q}$, and $\mathcal{Q}$ is a predefined, bounded set containing the origin.

Secondly, as the set $\mathcal{Q}$ is not known to be positive invariant, following \cref{def:LocalLyapunovFunction}, a positive invariant set $\mathcal{X} \subseteq \mathcal{Q}$ is required for the function $V$ satisfying \cref{eq:DecreaseConditionInformationTerm}.
By \cref{eq:DecreaseConditionInformationTerm}, a potential trivial class of invariant sets in $\mathcal{Q}$ consists of sublevel sets of $V$, $\mathcal{L}_\gamma(V) \coloneq \{ {x} \in \realsN{n} \mid V\left(x\right) \leq \gamma \}$. Therefore, a second \acrshort{SOS} optimization problem attempts to find the largest sublevel set of $V$ in $\mathcal{Q}$ \cite{mybibfile:Korda2017}.

Once a non-negative function $V$ satisfying \cref{eq:DecreaseConditionInformationTerm} and a positive invariant set $\mathcal{X} \subseteq \mathcal{Q}$ have been found, it follows that $V$ is a local Lyapunov function in $\mathcal{X}$ for $x^+ = f\big(x, \varphi(D_x^{-1}x)\big)$ \cite{mybibfile:Detailleur2025}, thereby certifying the closed-loop system is \acrshort{LAS} and providing an inner estimate of its \acrshort{ROA}.

\subsection{\texorpdfstring{\acrshort{SOS}-Compatible System Description}{SOS-Compatible System Description}}
\label{sec:StabilityVerification_SystemDescription}
To set up the \acrshort{SOS} optimization problems verifying the closed-loop stability properties, first, the closed-loop system $f\big(x,\varphi(D_x^{-1} x)\big)$ is expressed in the scaled coordinates used during the training of the \acrshort{NNC}, leading to the equivalent system
\begin{equation}
    \label{eq:ScaledSigiDTLTImodel}
    \bar{x}^+ = \bar{f}\big(\bar{x}, \varphi(\bar{x})\big) = \bar{A}\bar{x} + \bar{B}\varphi(\bar{x}),
\end{equation}
where $\bar{A} = D_x^{-1} A D_x$ and $\bar{B} = D_x^{-1}B$.

Next, the closed set $\bar{\mathcal{Q}}$ is defined as the region over which the (decrease) condition \cref{eq:DecreaseConditionInformationTerm} must hold. It contains the origin in its interior and is formally described by
\begin{equation}
    \bar{\mathcal{Q}} = \big\{ \bar{x} \in \realsN{n} \mid \bar{q}(\bar{x}) \geq 0 \big\},
    \label{eq:LocalRegionQDefinition}
\end{equation}
where $\bar{q}\colon \realsN{n} \mapsto \realsN{}$ is a quadratic function, $\bar{q}(\bar{x}) = \alpha - \bar{x}\transpose \bar{Q} \bar{x}$. 

\begin{figure*}[t]
    \begin{subfigure}{0.65\textwidth}
        \centering
        \def\svgwidth{0.95\linewidth}
        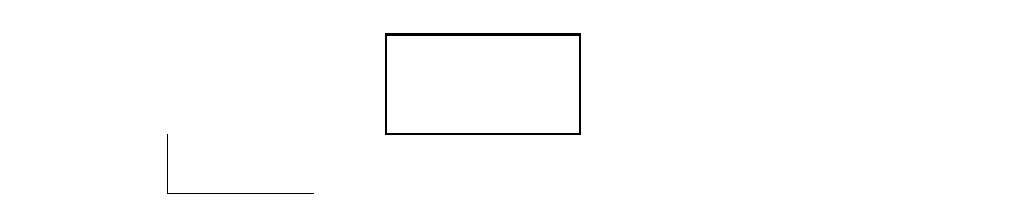
        \caption{}
        \label{subfig:OpenLoopGraph}
    \end{subfigure}
    \hfill
    \centering
    \begin{subfigure}{0.33\textwidth} 
        \centering    
        \begin{tikzpicture}
  \begin{axis}[
      width=4.5cm,
      height=2.5cm,
      scale only axis,
      axis lines=middle,
      xmin=-4.5, xmax=4.5,
      ymin=-1.5, ymax=3.5,
      xtick={-4,-3,...,4},
      ytick={-1,0,...,4},
      grid=major,
      grid style={gray!30},
      xlabel={$x$},
      ylabel={$\color{blue} \mathrm{ReLU}(x)$},
      xlabel style={
        at={(axis cs:3.75,0.125)},
        anchor=south west,
        font=\footnotesize,
      },
      ylabel style={
        at={(axis cs:0.125,2.5)},
        anchor=south west,
        font=\footnotesize,
        fill=white,
        fill opacity=0.9,
        inner sep=2pt,
      },
      tick label style={
        font=\scriptsize,
      },
    ]

    \addplot [
      pattern=north east lines,
      pattern color=gray,
      draw=none,
      fill opacity=0.75,
    ] coordinates {
      (-4.5,0)
      (4.5,0)
      (4.5,-1.5)
      (-4.5,-1.5)
      (-4.5,0)
    };

    \addplot [
      fill=gray,
      fill opacity=0.25,   
      draw=none,
    ] coordinates {
      (-4.5,0)
      (4.5,0)
      (4.5,-1.5)
      (-4.5,-1.5)
      (-4.5,0)
    };

    \addplot [
      pattern=north west lines,
      pattern color=gray,
      draw=none,
      fill opacity=0.75,
    ] coordinates {
      (-1.5,-1.5)
      (3.5,3.5)
      (4.5,3.5)
      (4.5,-1.5)
      (-1.5,-1.5)
    };

    \addplot [
      fill=gray,
      fill opacity=0.25,  
      draw=none,
    ] coordinates {
      (-1.5,-1.5)
      (3.5,3.5)
      (4.5,3.5)
      (4.5,-1.5)
      (-1.5,-1.5)
    };

     \draw[-Latex, thick, black] (axis cs:-2.5,-0.5) -- (axis cs:-2.5,0.75) node[above, font=\footnotesize] {};

     \draw[-Latex, thick, black] (axis cs:2.353,1.647) -- (axis cs:1.5,2.5) {};

    \addplot [thick, gray, domain=-1.5:0] {x};

    \addplot[very thick, blue] {max(0,x)};

    \addplot [
      red,
      very thick,
      dashed,
      dash pattern=on 2pt off 5pt,
      domain=-4.5:4.5,
      samples=2, 
    ] {x};
    
    \addplot [
      red,
      very thick,
      dashed,
      dash pattern=on 2pt off 5pt,
      domain=-4.5:4.5,
      samples=2,
    ] {0};

    \draw[black, thick] (rel axis cs:0,0) rectangle (rel axis cs:1,1);
  \end{axis}
\end{tikzpicture}
        \caption{}
        \label{subfig:NeuralNetworkGraph}
     \end{subfigure} 
    \caption{\subref{subfig:OpenLoopGraph} A block diagram of the \acrshort{NNC} $\varphi(\bar{x})$ and composed loop $L(\bar{x}) = \varphi \circ \bar{f} \circ (\textrm{id}, \varphi)$ making up the system model. \subref{subfig:NeuralNetworkGraph} A visualization of the graph of a \acrshort{ReLU} neuron that is defined by two inequality constraints, \acrshort{ReLU}$(x) \geq x$ and \acrshort{ReLU}$(x) \geq 0$, indicated by the two arrows, and one equality constraint, \acrshort{ReLU}$(x)($\acrshort{ReLU}$(x)-x)=0$, indicated by the red dashed line.}
    \label{fig:SemialgebraicSetModel}
\end{figure*}

To construct the polynomial term $\Gamma\big(\bar{x}, \, \bar{x}^+ ; (\bar{f}, \varphi, \bar{\mathcal{Q}})\big)$ of \cref{eq:DecreaseConditionInformationTerm}, \acrshort{SOS}-compatible input-output relations of the \acrshort{NNC} $\varphi$ 
and the composed loop $L = \varphi \circ \bar{f} \circ (\textrm{id}, \varphi)$
for all $\bar{x} \in \bar{\mathcal{Q}}$ are required. These relations, shown in \cref{subfig:OpenLoopGraph},  are obtained by means of their graphs \cite{mybibfile:Korda2022,mybibfile:Newton2022}.
\begin{defn}[Graph of a function]
    Given a function $f:\mathcal{S}\rightarrow \mathcal{V}$, its graph is defined by the set
    \vspace{-0.3\baselineskip}
    \begin{equation}
        \big\{ \big(x,y\big) 
        \; \big\vert \; x \in \mathcal{S}, \, y = f(x)  \big\}.
        \label{eq:GraphDefinition}
    \end{equation}
\end{defn}
\vspace{0.3\baselineskip}

The input-output relations of $\varphi$ and $L$ are found by first considering the graph of a single \acrshort{ReLU} neuron, $y = \max(0, \, w\transpose x + b)$, for all $x \in \realsN{n}$, which is described by
\begin{equation}
    \label{eq:SingleReLUNetwork}
    \left\{ (x, y) 
    \, \left\vert \:
    \begin{alignedat}{1}
        & y \geq 0, \ y - w\transpose x - b \geq 0 \\
        & y \big(y - w\transpose x - b \big) = 0
    \end{alignedat}
    \right. \right\}
\end{equation}
and shown visually in \cref{subfig:NeuralNetworkGraph}. From \cref{eq:OverallFFNeuralNet} it follows that a description of \acrshort{NNC} $\varphi$ can be obtained by composing the single-neuron graph description across successive layers of the network. Thus, introduce lifting variable $\lambda\transpose = [\lambda_1\transpose \, , \ldots , \, \lambda_\ell\transpose]\transpose \in \realsN{n_\lambda}$ to represent the output of all (\acrshort{ReLU}) neurons comprising \acrshort{NNC} $\varphi$,  
\begin{equation}
       \lambda_0 = \bar{x}, \ \lambda_{i} = \phi_i \circ f^{\varphi}_i \, (\lambda_{i-1})  \quad \forall i \in \left[\ell\right].
    \label{eq:HiddenLayerNeuralNet}
\end{equation}
Application of \cref{eq:SingleReLUNetwork} to all neurons in the \acrshort{NNC} $\varphi$ leads to the sets
\begin{equation}
    \label{eq:RecursiveReLUSemialgebraicSet}
    \left\{ (\lambda_{i-1}, \lambda_i) 
    \, \left\vert \:
    \begin{alignedat}{1}
        & \lambda_i \geq 0, \ \lambda_i - W_i \lambda_{i-1} - b_i \geq 0 
        \\ & \lambda_i \odot \left(\lambda_i - W_i \lambda_{i-1} - b_i \right) = 0
    \end{alignedat}
    \right. \right\},
\end{equation}
for all $i \in [\ell]$. Taking the union of i) all sets defined by \cref{eq:RecursiveReLUSemialgebraicSet} and ii) $n_u$ additional equality constraints defining $\varphi(\bar{x})$ as an affine transformation of $\lambda_\ell$, the graph of $\bar{\mathcal{Q}} \ni \bar{x} \mapsto [\lambda(\bar{x})\transpose, \, \varphi(\bar{x})\transpose]\transpose$, which includes an exact description of the input-output relationship of \acrshort{NNC} $\varphi$ for all $\bar{x} \in \bar{\mathcal{Q}}$, is given by
\begin{equation}
    \label{eq:SemialgebraicNetworkSet}
    \begin{aligned}
         \mathbf{K}_\varphi = \Bigg\{ \bigg(\bar{x}, \begin{bmatrix} \lambda \\ \varphi \end{bmatrix} \bigg) 
         \;  \left\vert \; \begin{bmatrix}
                g(\bar{x}, \lambda, \varphi) \\
                \bar{q}(\bar{x})
            \end{bmatrix} \geq 0, \right.  
            h(\bar{x}, \lambda, \varphi) = 0  \Bigg\}.
    \end{aligned}
\end{equation}
Next, an exact input-output description of the composed loop $L$ is constructed by additionally composing the scaled control-oriented model of \cref{eq:ScaledSigiDTLTImodel}. This allows the graph of $\bar{\mathcal{Q}} \ni \bar{x} \mapsto [\lambda(\bar{x})\transpose, \, \allowdisplaybreaks \varphi(\bar{x})\transpose, \, \allowdisplaybreaks {\bar{x}^+}(\bar{x})\transpose, \, \allowdisplaybreaks \lambda^+(\bar{x}), \, \allowdisplaybreaks \varphi^+(\bar{x})\transpose]\transpose$ to be expressed as
\begin{equation}
    \label{eq:SemialgebraicComposedLoopSet}  
    \begin{aligned}
         & \mathbf{K}_L =  \left\{ 
        \Big(\bar{x}, \, [\lambda\transpose\!\!, \, \varphi\transpose\!\!, \, ({\bar{x}}^{+})\transpose\!\!\!\!, \,
        (\lambda^+)\transpose\!\!\!\!, \, ({\varphi^+})\transpose]\transpose\Big) 
        \;  \left\vert
        \vphantom{\begin{bmatrix}
    		g(\bar{x}, \lambda, \varphi) \\
            \bar{q}(\bar{x}) \\
            g(\bar{x}^+, \lambda^+, \varphi^+)
        \end{bmatrix}} \right. \right. \\  & \qquad  \underbrace{\begin{bmatrix}
    		g(\bar{x}, \lambda, \varphi) \\
            \bar{q}(\bar{x}) \\
            g(\bar{x}^+, \lambda^+, \varphi^+)
        \end{bmatrix}}_{g^L} \geq 0, \, 
        \underbrace{\begin{bmatrix}
    		h(\bar{x}, \lambda, \varphi) \\
            h(\bar{x}^+, \lambda^+, \varphi^+)  \\
    		\bar{x}^+ - \bar{f}(\bar{x}, \varphi)  
        \end{bmatrix}}_{h^L} = 0 \left. \vphantom{\begin{bmatrix}
    		g(\bar{x}, \lambda, \varphi) \\
            \bar{q}(\bar{x}) \\
            g(\bar{x}^+, \lambda^+, \varphi^+)
        \end{bmatrix}} \!
        \right\}.
    \end{aligned}
\end{equation}
Thus, the sets described by \cref{eq:SemialgebraicNetworkSet,eq:SemialgebraicComposedLoopSet} \textit{exactly} define the input-output relations of \acrshort{NNC} $\varphi$ and composed loop $L$ for all $\bar{x} \in \bar{\mathcal{Q}}$. Furthermore, these sets are semialgebraic, i.e. consisting of polynomial (in)equalities, thereby rendering them compatible with \acrshort{SOS} programming/\acrshortpl{SDP}. 

For brevity of notation in the subsequent steps, following \cref{eq:SemialgebraicNetworkSet,eq:SemialgebraicComposedLoopSet}, let $\zeta(\bar{x})\transpose$ denote $[\bar{x}\transpose, \, \lambda(\bar{x})\transpose, \, \varphi(\bar{x})\transpose]\transpose$, and let $\xi(\bar{x})\transpose$ denote $[\zeta(\bar{x})\transpose, \, \zeta^+(\bar{x})\transpose]\transpose$, which we note are both continuous functions in $\bar{x}$ by \cref{eq:SigiDTLTImodel,eq:OverallFFNeuralNet,eq:ScaledSigiDTLTImodel}. 

\subsection{Search For Strictly Decreasing Function V}
Using the \acrshort{SOS}-compatible system description of $\mathbf{K}_\varphi$ and $\mathbf{K}_L$, a strictly decreasing function $V$ is now obtained via the use of \acrshort{SOS} polynomials.

\begin{defn}[\Acrfull{SOS} polynomial]
    A polynomial $\sigma$ is \acrshort{SOS} if it admits a decomposition as a sum of squared polynomials. Given a vector of appropriate monomial terms $\nu(\zeta)$ and a matrix of coefficients $L$, \acrshort{SOS} polynomials admit an equivalent semidefinite representation 
    \begin{equation}
        \label{eq:SOSDefinition}
        \sigma(\zeta) = \sum_i \sigma_i(\zeta)^2 = \sum_i (l_i\transpose \nu(\zeta))^2 = \nu(\zeta)\transpose \underbrace{L\transpose L}_{\succeq 0} \nu(\zeta),        
    \end{equation}    
    This relation allows \acrshort{SOS} polynomials to be used in \acrshortpl{SDP}. See the work of Parrilo \cite{mybibfile:Parrilo2003} for more information.
\end{defn}

Using this definition and the semialgebraic set $\mathbf{K}_\varphi$, 
a rich class of non-negative polynomial functions $V:\mathbb{R}^{n}\rightarrow \mathbb{R}_{\geq 0}$ in $\bar{x}$ are parameterized via
\begin{equation}
    \label{eq:LyapunovParam}
    V(\zeta) = \sigma^V(\zeta) + \sigma^V_{\textrm{ineq}}(\zeta)\transpose 
    \underbrace{\begin{bmatrix}
        \mathcal{M}\big(g(\zeta), 1\big) \\ 
        \mathcal{M}(g(\zeta), 2\big) \\ 
        \vdots
    \end{bmatrix}}_{g^V},
\end{equation}
with parameters $\sigma^V$, $\sigma^V_{\textrm{ineq}}$  representing any scalar \acrshort{SOS} polynomial and  any vector of \acrshort{SOS} polynomials, respectively. Here $\zeta$ is understood to be a function of $\bar{x}$. Note how the functions parameterized by \cref{eq:LyapunovParam} are 
not necessarily candidate Lyapunov functions satisfying \cref{eq:V_conditions_a,eq:V_conditions_b}, 
as they are not all equal to $0$  for $\bar{x} = 0$.

Following the form of \cref{eq:DecreaseConditionInformationTerm} and using $\mathbf{K}_L$ to construct $\Gamma\big(\bar{x}, \, \bar{x}^+ ; (\bar{f}, \varphi, \bar{\mathcal{Q}})\big)$, a sufficient condition to ensure the value of $V$ decreases over time for all $\bar{x} \in \bar{\mathcal{Q}}$ is formulated as
\begin{multline}
    V(\zeta) - V(\zeta^+) - \|\bar{x}\|_P^2  -p^{\Delta V}_{\textrm{eq}}(\xi)\transpose h^L(\xi) \\  - \sigma^{\Delta V}(\xi)- \sigma^{\Delta V}_{\textrm{ineq}}(\xi) \transpose
    \begin{bmatrix}
        \mathcal{M}\big(g^L(\xi), 1\big) \\ 
        \mathcal{M}\big(g^L(\xi), 2\big)\\ 
        \vdots
    \end{bmatrix} \geq 0
    \label{eq:LocalLyapunovDecreaseCond}
\end{multline}
with $P$ any positive definite matrix, $p^{\Delta V}_{\textrm{eq}}$ any vector of arbitrary polynomials, $\sigma^{\Delta V}$ any scalar \acrshort{SOS} polynomial and $\sigma^{\Delta V}_{\textrm{ineq}}$ any vector of \acrshort{SOS} polynomials.

\Cref{eq:LocalLyapunovDecreaseCond} can be brought to the canonical form $\nu^{\Delta V}_{\textrm{tot}}(\xi)\transpose Q^{\Delta V}_{\textrm{tot}} \nu^{\Delta V}_{\textrm{tot}}(\xi) \geq 0$, where $\nu^{\Delta V}_{\textrm{tot}}(\xi)$ represents a basis vector (of monomials). Then, proving $Q^{\Delta V}_{\textrm{tot}} \succeq 0$ establishes that the left-hand side of \cref{eq:LocalLyapunovDecreaseCond} is an \acrshort{SOS} polynomial, which is a sufficient condition to prove the inequality. 
Using this well-established relation between \acrshort{SOS} polynomials and quadratic forms allows a function $V$ satisfying \cref{eq:LocalLyapunovDecreaseCond} to be found by solving the \acrshort{SDP}
\begin{subequations}
    \label{eq:SDPFormulationLocalAsymptoticStability}
    \begin{alignat}{4}
        &\span\span \text{find:} \ & \sigma^V, \, \sigma^V_{\textrm{ineq}}, \, P, \ \,
        & 
        \! \! \sigma^{\Delta V},  \, \sigma^{\Delta V}_{\textrm{ineq}}, \,  p^{\Delta V}_{\textrm{eq}} \span \nonumber \\
        &\span\span \text{s.t.} \ & \eqref{eq:LyapunovParam}, \ \, & \! \! \eqref{eq:LocalLyapunovDecreaseCond}, \span \\
        &\span\span & \sigma^V, \, \sigma^V_{\textrm{ineq}}, \sigma^{\Delta V},  \, \sigma^{\Delta V}_{\textrm{ineq}} \quad & \text{\acrshort{SOS} polynomials,} \label{eq:LocalAsymptoticStabilitySOS} \\
        &\span\span & 
        p^{\Delta V}_{\textrm{eq}} \quad & \text{arbitrary polynomials.} \label{eq:LocalAsymptoticStabilityP} \\
         &\span\span & P \succ & \, \, 0.  \label{eq:InvariantLocalAsymptoticStabilityGammaStrictlyPositive}
    \end{alignat}
\end{subequations}

\subsection{\texorpdfstring{Search For Largest Sublevel Set in $\bar{\mathcal{Q}}$}{Search For Largest Sublevel Set in Q}}
Subsequently, to obtain an inner estimate of the closed-loop system's \acrshort{ROA} and certify that the function $V$ returned by \acrshort{SDP} problem \cref{eq:SDPFormulationLocalAsymptoticStability} is a valid local Lyapunov function, the largest sublevel set of $V$ contained in the set $\bar{\mathcal{Q}}$ is sought. Consider the condition
\begin{multline}
    \sigma^{\bar{\mathcal{Q}}}_q(\zeta) \bar{q}(\bar{x}) \geq \big( \gamma - V(\zeta) \big) + p^{\bar{\mathcal{Q}}}_{\textrm{eq}}(\zeta)\transpose h(\zeta)\\ + \sigma^{\bar{\mathcal{Q}}}(\zeta)
     + \sigma^{\bar{\mathcal{Q}}}_{\textrm{ineq}}(\zeta)\transpose  
    \begin{bmatrix}
        \mathcal{M}\big(g(\zeta), 1\big) \\
        \mathcal{M}\big(g(\zeta), 2\big) \\ 
        \vdots
    \end{bmatrix},
    \label{eq:SublevelSetCond}
\end{multline}
with \smash{$\sigma^{\bar{\mathcal{Q}}}_q$}, $\sigma^{\bar{\mathcal{Q}}}$ any scalar \acrshort{SOS} polynomials, $p^{\bar{\mathcal{Q}}}_{\textrm{eq}}$ any vector of arbitrary polynomials and \smash{$\sigma^{\bar{\mathcal{Q}}}_{\textrm{ineq}}$} any vector of \acrshort{SOS} polynomials. 
Note that if \cref{eq:SublevelSetCond} is satisfied it holds that for any $\bar{x}$ such that $V(\zeta(\bar{x})) \leq \gamma$, $q(\bar{x}) \geq 0$. Thus, verifying \cref{eq:SublevelSetCond} certifies that $\mathcal{L}_\gamma(V) \subseteq \mathcal{Q}$.
In a manner similar to \acrshort{SDP} problem \cref{eq:SDPFormulationLocalAsymptoticStability}, the condition of \cref{eq:SublevelSetCond} can be used to formulate the optimization problem
\begin{subequations}
    \label{eq:SDPFormulationLargestSublevelSet}
    \begin{alignat}{4}
        &\span\span \underset{\sigma^{\bar{\mathcal{Q}}}_q, \, \sigma^{\bar{\mathcal{Q}}}, \, \sigma^{\bar{\mathcal{Q}}}_{\textrm{ineq}}, \, \gamma}{\textrm{maximize:}} \ &   & \! \! \! \! \! \! \! \! \gamma  \span \nonumber \\
        &\span\span \text{s.t.} \ & \eqref{eq:LyapunovParam}, \ \, & \! \! \eqref{eq:SublevelSetCond}, \span \\
        &\span\span & \sigma^{\bar{\mathcal{Q}}}_q, \, \sigma^{\bar{\mathcal{Q}}}, \, \sigma^{\bar{\mathcal{Q}}}_{\textrm{ineq}} \quad & \text{\acrshort{SOS} polynomials,} \label{eq:test_2} \\
        &\span\span & p^{\bar{\mathcal{Q}}}_{\textrm{eq}} \quad & \text{arbitrary polynomials,} \label{eq:test_3}
    \end{alignat}
\end{subequations}
which searches for the largest sublevel set $\mathcal{L}_\gamma(V) \subseteq \bar{\mathcal{Q}}$. 

If optimization problems \cref{eq:SDPFormulationLocalAsymptoticStability,eq:SDPFormulationLargestSublevelSet} are solved in succession, it follows that $V$ is a local Lyapunov function in $\mathcal{L}_\gamma(V)$ for $\bar{x}^+ = \bar{f}\big(\bar{x}, \varphi(\bar{x})\big)$, proving the closed-loop system is \acrshort{LAS} and providing an inner estimate of its \acrshort{ROA}.

\subsection{Implementation and Practical Considerations}
\label{sec:ImplementationAndPracticalConsiderations}
To find a stability certificate using optimization problems \cref{eq:SDPFormulationLocalAsymptoticStability,eq:SDPFormulationLargestSublevelSet} for the \acrshort{NNC} synthesized in \cref{sec:ControllerSynthesis_NNC}, second-order products of $g$ and $g^L$ are considered. 

Next, the basis vectors that form the \acrshort{SOS} polynomials in all optimization problems must be fixed. Following the derivation of \cref{sec:StabilityVerification_SystemDescription}, it follows that at least one of the inequalities of \cref{eq:SingleReLUNetwork} is tight for all $\bar{x}$. Thus, let the index set $\mathcal{I}_0$ represent the entries of $g(\zeta)$ equal to zero for $\bar{x} = 0$, and let $\mathcal{I}^{c}_0$ denote its complement. The basis vectors of $\sigma^V$, $\sigma^{\Delta V}$, $\sigma^{\bar{\mathcal{Q}}}$ are then fixed according to \cref{tab:SOSmultipliers}.

To minimize the size of the optimization problem, the \acrshort{SOS} polynomials in \cref{eq:LyapunovParam} are chosen such that they have a minimum at $\bar{x} = 0$  \cite{mybibfile:Detailleur2025}. This is done by selecting the entries of $\sigma^V_{\textrm{ineq}}$ according to \cref{tab:SOSmultipliers}.

Under the condition that no terms in \cref{eq:LocalLyapunovDecreaseCond}, \cref{eq:SublevelSetCond} exceed the degree of those generated by $\sigma^{\Delta V}$, $\sigma^{\bar{\mathcal{Q}}}$, respectively, the remaining polynomial terms are chosen to be of maximal size. 

Finally, the set $\bar{\mathcal{Q}}$ to be examined must be specified. The matrix $\bar{Q}$ defining this set is determined by examining the linearization of the closed-loop system at $\bar{x}=0$ and solving
\begin{equation}
    \label{eq:LocalLinearizationLyapunovEquation}
    \bigg(\bar{A} + \bar{B}\od{\varphi(\bar{x})}{\bar{x}}\Big\vert_{\bar{x}=0}\bigg)\transpose\bar{Q}\bigg(\bar{A} + \bar{B}\od{\varphi(\bar{x})}{\bar{x}}\Big\vert_{\bar{x}=0}\bigg) - \bar{Q} = -I.
\end{equation} 
With optimization problems \cref{eq:SDPFormulationLocalAsymptoticStability,eq:SDPFormulationLargestSublevelSet} specified up to the value of $\alpha$ in the definition of $\bar{q}(\bar{x})$, these optimization problems are solved using SOSTOOLS \cite{mybibfile:SOSTOOLS} and MOSEK \cite{mybibfile:MOSEK} for increasing values of $\alpha$, until optimization problem \cref{eq:SDPFormulationLocalAsymptoticStability} is no longer feasible.

\begin{table}[t]
\centering
\begin{tabular}{ll}
\hline
\multicolumn{1}{c}{\shortstack[c]{\textbf{SOS polynomial}${}^{\phantom{|}}$ \\ $\sigma^{(...)} = \nu\transpose P^{(...)} \nu$}}
    & \multicolumn{1}{c}{\shortstack[c]{\textbf{\phantom{y}Basis vector\phantom{y}}${}^{\phantom{|}}$ \\ \phantom{${\sigma^{(...)}}$}$\nu$\phantom{$\sigma^{(...)}$}}}
    \\ \hline
${\sigma^V}^{\phantom{|}}$                 & $\phantom{=} \quad \: [\bar{x}\transpose, 
\, g_{\mathcal{I}_0}(\zeta)\transpose, \, \varphi\transpose, \, \mathcal{M}([\bar{x}\transpose, \, \varphi\transpose]\transpose, \,2)\transpose]\transpose$                              \\
$\sigma^{\Delta V}$ & 
$\begin{matrix}
\phantom{=} \quad \: [1,\, \bar{x}\transpose,\, g_{\mathcal{I}_0}(\zeta)\transpose,\, \varphi\transpose, \, (\bar{x}^+)\transpose,\, g_{\mathcal{I}_0}(\zeta^+)\transpose, \, \dots \qquad \qquad \\ \qquad \qquad (u^+)\transpose, \, \mathcal{M}([\bar{x}\transpose,\, (\bar{x}^+)\transpose,\, \varphi\transpose,\, (\varphi^+)\transpose]\transpose,\,2)\transpose]\transpose \end{matrix} $ \\
$\sigma^{\mathcal{\bar{Q}}}$     & $\phantom{=} \quad \: [1, \, \bar{x}\transpose, g_{\mathcal{I}_0}(\zeta)\transpose, \, \varphi\transpose, \, \mathcal{M}([\bar{x}\transpose, \, \varphi\transpose]\transpose, \,2)\transpose]\transpose$   \\
$\sigma^V_{\textrm{ineq},i}$ & $ =
\begin{cases}
[1, \bar{x}\transpose, \, \varphi\transpose]\transpose & \text{if } g^V_i \in \{g_i\}_{i \in {\mathcal{I}_0}}, \\
[\bar{x}\transpose, \, \varphi\transpose]\transpose & \text{if } g^V_i \in \{g_i\}_{i \in  \mathcal{I}^c_0}, \\
1 & \text{if } g^V_i \in \{g_ig_j\}_{ i \in [\textrm{dim}(g)], \,  j\in \mathcal{I}_0 } 
\\
0 & \text{otherwise}.
\end{cases}$                                
\end{tabular} 
\caption{An overview of the basis vectors used in the \acrshort{SOS} optimization problems.}
\label{tab:SOSmultipliers}
\end{table}

\begin{figure*}[t]
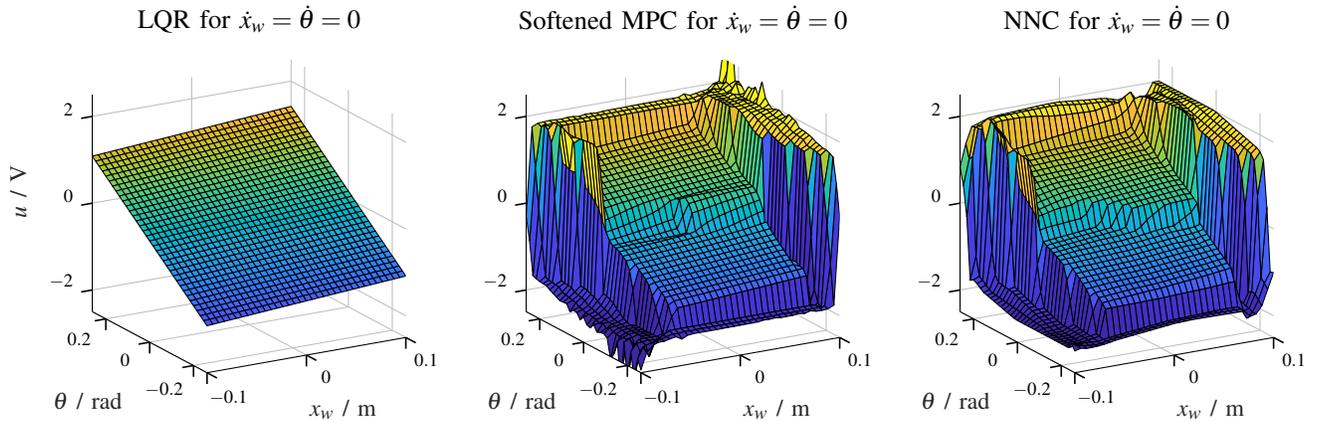

	\begin{center}
		%
%
	\end{center}
	\caption{A graphical comparison between the control laws defined by the \acrshort{LQR}, \acrshort{MPC} and \acrshort{NNC} for $\dot{x}_w = \dot{\theta} = 0$.}
	    \label{fig:ControlLaw}
\end{figure*}

\section{Results \& Discussion}
\label{sec:Results}
This section presents and discusses the results that demonstrate the value of the generalizable \acrshort{NNC} synthesis procedure and the \acrshort{SOS}-based stability verification procedure for controlling the two-wheeled inverted pendulum system.

First, a qualitative assessment of the \acrshort{NNC} is presented via a graphical comparison between the control outputs of the \acrshort{LQR}, robust, tube-based \acrshort{MPC} and \acrshort{NNC} over the $(x_w, \theta)$ subspace. Next, the properties of the \acrshort{NNC}'s local stability certificate obtained via the procedure of \cref{sec:StabilityVerification} are examined. Finally, experimental validation on the hardware is performed, whereby the control performance of the baseline \acrshort{LQR} and the \acrshort{NNC} are compared in both a regulation and a reference-tracking task. 

\subsection{Qualitative Assessment of the Synthesized NNC}
\label{sec:Results_QualitativeComparison}
A qualitative comparison of the three controllers synthesized in \cref{sec:ControllerSynthesis} is achieved by means of a surface plot of each control law. \Cref{fig:ControlLaw} shows the control inputs of the \acrshort{LQR}, softened \acrshort{MPC} and \acrshort{NNC} over the $(x_w, \theta)$ subspace of the four-dimensional state space.

As expected from optimization problem \cref{eq:RobustMPC_OptimizationProblem}, the softened, robust, tube-based \acrshort{MPC} exhibits significant nonlinearities near the origin and close to the border of allowable states, making it substantially more complex than the \acrshort{LQR}. In addition, \cref{fig:ControlLaw} shows how, with help of the sampling and constraint softening steps of \cref{sec:ControllerSynthesis_NNC}, the \acrshort{NNC} also exhibits these nonlinearities. This qualitative similarity indicates that the generalizable synthesis procedure allows the \acrshort{NNC} to successfully approximate the complex and highly nonlinear behavior of the MPC. Furthermore, the additional neurons added to saturate the \acrshort{NNC} output can be seen at $(x_w, \theta) \in \{ (0.1, \, 0.2), (-0.1, \, -0.2) \}$.

\subsection{Local Stability Properties of NNC}
By application of the procedure described in \cref{sec:StabilityVerification}, a local stability certificate in the form of a local Lyapunov function is obtained for the closed-loop system using the \acrshort{NNC} of \cref{sec:ControllerSynthesis_NNC}. 

Using the trained and postprocessed \acrshort{NNC}, \cref{eq:LocalLinearizationLyapunovEquation} is solved, yielding
\begin{equation}
    \bar{Q} = \begingroup
        \sisetup{print-exponent-implicit-plus=true}
        \sisetup{tight-spacing=true}
        \setlength{\arraycolsep}{2pt}
        \begin{bNiceMatrix}
            \num{9.635753395702750e+01}     & \num{8.352818772147614e+01}     & \num{1.054573377306265e+01}     & \num{5.375827694114691e+01}     \\
            *                               & \num{1.867815684474361e+02}     & \num{1.830758126302962e+01}     & \num{1.200136008099704e+02}     \\
            *                               & *                          & \num{7.332093812671625e+00}e\!+\!0  & \num{1.238447787535704e+01}       \\
            *                               & *                          & *                            & \num{7.863364772994072e+01}       \\
        \end{bNiceMatrix}
    \endgroup.
\end{equation}

Using the resulting set $\bar{\mathcal{Q}}$, optimization problem \cref{eq:SDPFormulationLocalAsymptoticStability} is solved and found feasible for values of $\alpha$ up to $1.5$. Given $V$ defined via the solution to \cref{eq:SDPFormulationLocalAsymptoticStability} with $\alpha$ set to $1.5$, the solution of optimization problem \cref{eq:SDPFormulationLargestSublevelSet} indicates that the sublevel set of $V$ corresponding to a value of $\gamma = 1.95$ lies inside $\bar{\mathcal{Q}}$. These results certify that the closed-loop system is \acrshort{LAS} and provide us with an inner estimate of its \acrshort{ROA}.

\begin{figure*}[t]
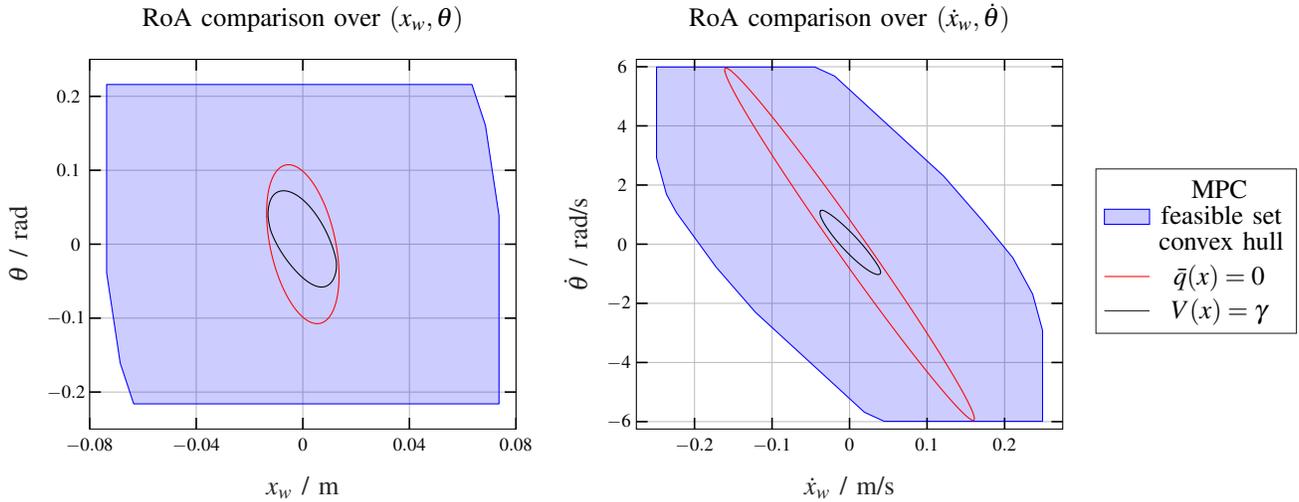

    \begin{center}
%
    \end{center}
    \caption{A comparison of i) a convex hull-based approximation of the (unsoftened) robust, tube-based \acrshort{MPC}'s feasible set, ii) the predefined set $\bar{\mathcal{Q}}$ of \cref{eq:LocalRegionQDefinition}, and iii) the proven inner estimate of the closed-loop system under the \acrshort{NNC}, $\mathcal{L}_\gamma(V)$.}
    \label{fig:RoA2MPCComparison}
\end{figure*}

Given the number of terms comprising the local Lyapunov function $V$, its value cannot be given in this work. 
However, the inner approximation of the \acrshort{ROA} this function defines is shown in \cref{fig:RoAComparison,fig:RoA2MPCComparison}. 

In each of the surface plots of \cref{fig:RoAComparison}, a single state variable is constrained to be equal to zero. This allows the $\gamma$-level set to be visualized as a surface, which constitutes the border of the proven inner approximation of the closed-loop system's \acrshort{ROA}. Additional contour plots depicting several level sets are also shown under the assumption that an additional, relevant state variable is equal to zero. In addition, \cref{fig:RoA2MPCComparison} shows how the set $\bar{\mathcal{Q}}$,  the inner estimate of the closed-loop system's \acrshort{ROA} and an approximation of the feasible set of the (unsoftened) robust, tube-based \acrshort{MPC} compare in the $(x_w, \theta)$ and $(\dot{x}_w, \dot{\theta})$ subspaces. 

These results clearly indicate that the inner estimate of the \acrshort{ROA} under the \acrshort{NNC}, while still smaller than the feasible set of the underlying robust, tube-based \acrshort{MPC}, is large enough to offer significant practical value, particularly when taking into account that the \acrshort{MPC} is not capable of running in real time on the embedded hardware. Additionally, the right plot over the $(\dot{x}_w, \dot{\theta})$ space indicates that the value of $\alpha$ in optimization problem \cref{eq:SDPFormulationLocalAsymptoticStability} likely could not be increased further due to the set $\bar{\mathcal{Q}}$ extending to the limit of the \acrshort{MPC}'s feasible set. Improved inner estimates of the \acrshort{ROA} would therefore likely require \cref{eq:SDPFormulationLocalAsymptoticStability} to be solved for a different set $\bar{\mathcal{Q}}$. Suggested improvements to the stability verification procedure are discussed in the conclusion of \cref{sec:Conclusion}.

\definecolor{mycolor1}{rgb}{0.30100,0.74500,0.93300}%
\definecolor{mycolor2}{rgb}{0.24220,0.15040,0.66030}%
\definecolor{mycolor3}{rgb}{0.20060,0.48030,0.99060}%
\definecolor{mycolor4}{rgb}{0.07700,0.74680,0.72240}%
\definecolor{mycolor5}{rgb}{0.78290,0.75790,0.16080}%
\definecolor{mycolor6}{rgb}{0.97690,0.98390,0.08050}%

\begin{figure*}[t]
    \begin{center}
       \begin{tikzpicture}
			\begin{groupplot}
				[
				group style={group name=RoAComparisonGroupPlot, group size=2 by 1, vertical sep=0.25cm, horizontal sep=3cm
				},
				width=7.5cm, height=6.5cm,
				]
				\nextgroupplot
				[
                ticklabel style = {
                    /pgf/number format/fixed,
                    /pgf/number format/precision=3
                },
                scaled ticks = false,
                xmin=-1.25,
                xmax=1.25,
                xtick={-1, -0.5, 0, 0.5, 1},
                tick align=outside,
                xlabel style={font=\color{white!15!black}},
                xlabel={$\dot{\theta}$ / rad/s},
                ymin=-0.05,
                ymax=0.05,
                ytick={-0.05, -0.025, 0, 0.025, 0.05},
                ylabel style={font=\color{white!15!black}},
                ylabel={$\dot{x}_w$ / m/s},
                zmin=-0.015,
                zmax=0.015,
                zlabel style={font=\color{white!15!black}},
                zlabel={$x_w$ / m},
                title={\acrshort{ROA} under \acrshort{NNC} for $\theta = \qty[]{0}{\radian}$},
                view={40}{10},
                axis background/.style={fill=white},
                axis x line*=bottom,
                axis y line*=left,
                axis z line*=left,
                xmajorgrids,
                ymajorgrids,
                zmajorgrids
                ]
                
                \addplot3[area legend, patch, shader=flat, fill=mycolor1, draw=black, draw opacity=0.1, forget plot, patch table={Isosurface_NNC_verification_20250606T1539_20250611T1543_const_theta-1.tsv}]
                table[] {Isosurface_NNC_verification_20250606T1539_20250611T1543_const_theta-2.tsv};
                
                \addplot3[area legend, draw=mycolor2, forget plot]
                table[] {Isosurface_NNC_verification_20250606T1539_20250611T1543_const_theta-3.tsv};
                
                \addplot3[area legend, draw=mycolor3, forget plot]
                table[] {Isosurface_NNC_verification_20250606T1539_20250611T1543_const_theta-4.tsv};
                
                \addplot3[area legend, draw=mycolor4, forget plot]
                table[] {Isosurface_NNC_verification_20250606T1539_20250611T1543_const_theta-5.tsv};
                
                \addplot3[area legend, draw=mycolor5, forget plot]
                table[] {Isosurface_NNC_verification_20250606T1539_20250611T1543_const_theta-6.tsv};
                
                \addplot3[area legend, draw=mycolor6, forget plot]
                table[] {Isosurface_NNC_verification_20250606T1539_20250611T1543_const_theta-7.tsv};
                
                \addplot3[area legend, draw=mycolor2, forget plot]
                table[] {Isosurface_NNC_verification_20250606T1539_20250611T1543_const_theta-8.tsv};
                
                \addplot3[area legend, draw=mycolor3, forget plot]
                table[] {Isosurface_NNC_verification_20250606T1539_20250611T1543_const_theta-9.tsv};
                
                \addplot3[area legend, draw=mycolor4, forget plot]
                table[] {Isosurface_NNC_verification_20250606T1539_20250611T1543_const_theta-10.tsv};
                
                \addplot3[area legend, draw=mycolor5, forget plot]
                table[] {Isosurface_NNC_verification_20250606T1539_20250611T1543_const_theta-11.tsv};
                
                \addplot3[area legend, draw=mycolor6, forget plot]
                table[] {Isosurface_NNC_verification_20250606T1539_20250611T1543_const_theta-12.tsv};
                
                \addplot3[area legend, draw=mycolor2, forget plot]
                table[] {Isosurface_NNC_verification_20250606T1539_20250611T1543_const_theta-13.tsv};
                
                \addplot3[area legend, draw=mycolor3, forget plot]
                table[] {Isosurface_NNC_verification_20250606T1539_20250611T1543_const_theta-14.tsv};
                
                \addplot3[area legend, draw=mycolor4, forget plot]
                table[] {Isosurface_NNC_verification_20250606T1539_20250611T1543_const_theta-15.tsv};
                
                \addplot3[area legend, draw=mycolor5, forget plot]
                table[] {Isosurface_NNC_verification_20250606T1539_20250611T1543_const_theta-16.tsv};
                
                \addplot3[area legend, draw=mycolor6, forget plot]
                table[] {Isosurface_NNC_verification_20250606T1539_20250611T1543_const_theta-17.tsv};
                \nextgroupplot
				[
                ticklabel style = {
                    /pgf/number format/fixed,
                    /pgf/number format/precision=3
                },
                scaled ticks = false,
                xmin=-1.25,
                xmax=1.25,
                xtick={-1, -0.5, 0, 0.5, 1},
                tick align=outside,
                xlabel style={font=\color{white!15!black}},
                xlabel={$\dot{\theta}$ / rad/s},
               ymin=-0.05,
                ymax=0.05,
                ytick={-0.05, -0.025, 0, 0.025, 0.05},
                ylabel style={font=\color{white!15!black}},
                ylabel={$\dot{x}_w$ / m/s},
                zmin=-0.1,
                zmax=0.1,
                zlabel style={font=\color{white!15!black}},
                zlabel={$\theta$ / rad},
                title={\acrshort{ROA} under \acrshort{NNC} for $\dot{x}_w = 0$ m/s},
                view={40}{10},
                axis background/.style={fill=white},
                axis x line*=bottom,
                axis y line*=left,
                axis z line*=left,
                xmajorgrids,
                ymajorgrids,
                zmajorgrids
                ]
                
                \addplot3[area legend, patch, shader=flat, fill=mycolor1, draw=black, draw opacity=0.1, forget plot, patch table={Isosurface_NNC_verification_20250606T1539_20250611T1543_const_x-1.tsv}]
                table[] {Isosurface_NNC_verification_20250606T1539_20250611T1543_const_x-2.tsv};
                
                \addplot3[area legend, draw=mycolor2, forget plot]
                table[] {Isosurface_NNC_verification_20250606T1539_20250611T1543_const_x-3.tsv};
                
                \addplot3[area legend, draw=mycolor3, forget plot]
                table[] {Isosurface_NNC_verification_20250606T1539_20250611T1543_const_x-4.tsv};
                
                \addplot3[area legend, draw=mycolor4, forget plot]
                table[] {Isosurface_NNC_verification_20250606T1539_20250611T1543_const_x-5.tsv};
                
                \addplot3[area legend, draw=mycolor5, forget plot]
                table[] {Isosurface_NNC_verification_20250606T1539_20250611T1543_const_x-6.tsv};
                
                \addplot3[area legend, draw=mycolor6, forget plot]
                table[] {Isosurface_NNC_verification_20250606T1539_20250611T1543_const_x-7.tsv};
                
                \addplot3[area legend, draw=mycolor2, forget plot]
                table[] {Isosurface_NNC_verification_20250606T1539_20250611T1543_const_x-8.tsv};
                
                \addplot3[area legend, draw=mycolor3, forget plot]
                table[] {Isosurface_NNC_verification_20250606T1539_20250611T1543_const_x-9.tsv};
                
                \addplot3[area legend, draw=mycolor4, forget plot]
                table[] {Isosurface_NNC_verification_20250606T1539_20250611T1543_const_x-10.tsv};
                
                \addplot3[area legend, draw=mycolor5, forget plot]
                table[] {Isosurface_NNC_verification_20250606T1539_20250611T1543_const_x-11.tsv};
                
                \addplot3[area legend, draw=mycolor6, forget plot]
                table[] {Isosurface_NNC_verification_20250606T1539_20250611T1543_const_x-12.tsv};
                
                \addplot3[area legend, draw=mycolor2, forget plot]
                table[] {Isosurface_NNC_verification_20250606T1539_20250611T1543_const_x-13.tsv};
                
                \addplot3[area legend, draw=mycolor3, forget plot]
                table[] {Isosurface_NNC_verification_20250606T1539_20250611T1543_const_x-14.tsv};
                
                \addplot3[area legend, draw=mycolor4, forget plot]
                table[] {Isosurface_NNC_verification_20250606T1539_20250611T1543_const_x-15.tsv};
                
                \addplot3[area legend, draw=mycolor5, forget plot]
                table[] {Isosurface_NNC_verification_20250606T1539_20250611T1543_const_x-16.tsv};
                
                \addplot3[area legend, draw=mycolor6, forget plot]
                table[] {Isosurface_NNC_verification_20250606T1539_20250611T1543_const_x-17.tsv};

                \end{groupplot}
		\end{tikzpicture}%
    \end{center}
    \caption{A visualization of the set certified to form part of the \acrshort{ROA} of the closed-loop system under the synthesized \acrshort{NNC}.}
    \label{fig:RoAComparison}
\end{figure*}
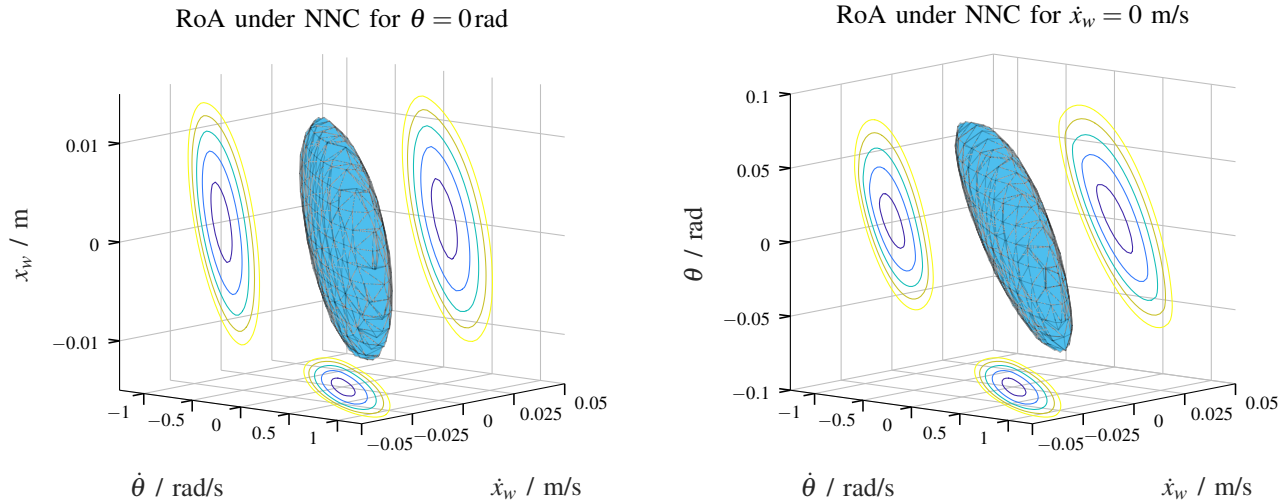

\subsection{Experimental Validation}
\definecolor{mycolor1}{rgb}{0.00000,0.44700,0.74100}%
\definecolor{mycolor2}{rgb}{0.85000,0.32500,0.09800}%
\definecolor{mycolor3}{rgb}{0.9290 0.6940 0.1250}%
\definecolor{mycolor4}{rgb}{0.49400,0.18400,0.55600}%

Finally, the empirical performance of both controllers capable of running in real time on the Sigi platform's embedded hardware are compared: the baseline \acrshort{LQR} and the \acrshort{NNC} with verified local stability properties. In all tests the state estimator of the Sigi platform is calibrated for $T_{\textrm{bias}} = \qty[scientific-notation=false,round-precision=2]{60}{\second}$ before a $\qty[scientific-notation=false,round-precision=2]{40}{\second}$ regulation or reference-tracking task is executed on the platform as described in \cref{sec:ControllerSynthesis_OutputModification}. Adjusting for the bias estimation by setting $t_0 = T_{\textrm{bias}}$, the reference-tracking task is defined by the parameters $v_x = \qty[scientific-notation=false,round-precision=1]{0.05}{\centi\metre\per\second}$, $t_1 = T_{\textrm{bias}} + \qty[scientific-notation=false,round-precision=2]{10}{\second}$,  $t_2 = T_{\textrm{bias}} + \qty[scientific-notation=false,round-precision=2]{30}{\second}$ and $t_3 = T_{\textrm{bias}} + \qty[scientific-notation=false,round-precision=2]{40}{\second}$. For each test, the initial state of the Sigi platform is set within the known \acrshort{ROA} of the \acrshort{NNC}. 

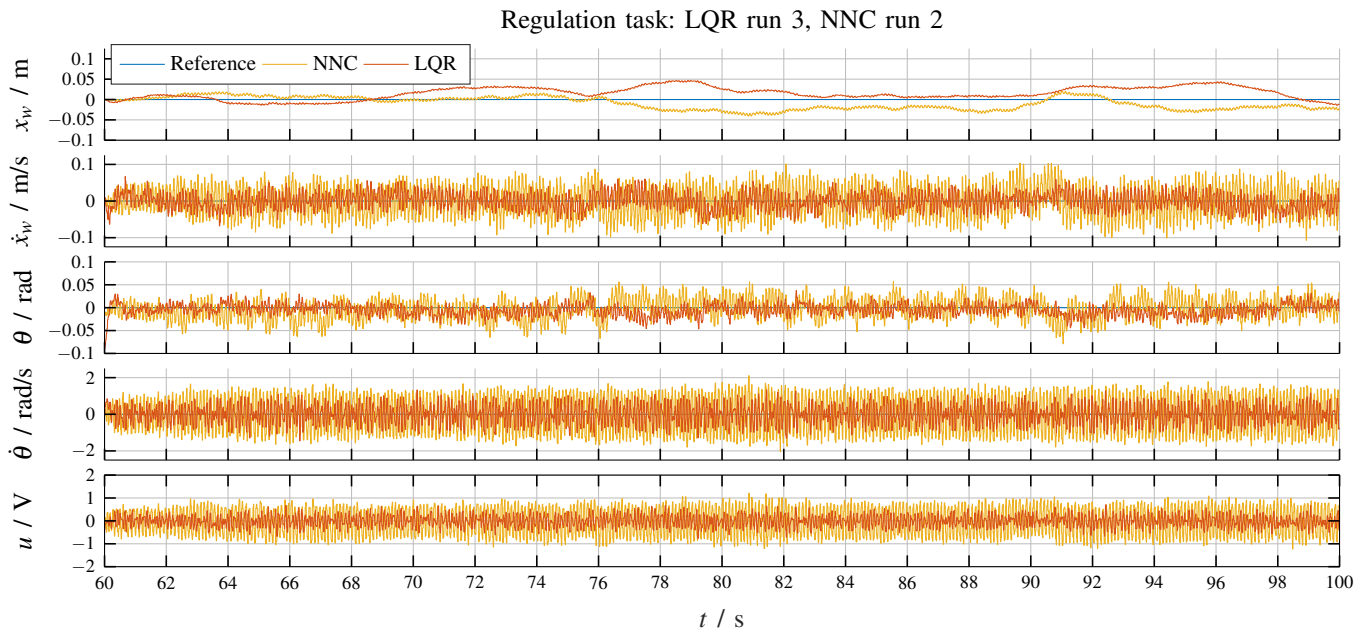
\begin{figure*}[t]
	\begin{center}
		\begin{tikzpicture}
			\begin{groupplot}
				[
				group style={group name=OriginTrackingGroupPlot, group size=1 by 5, vertical sep=0.2cm, horizontal sep=1cm
				},
				width=18cm, height= 2.8cm,
				]
				\nextgroupplot
				[
				xmin=60,
                xmax=100,
                ymin=-0.1,
                ymax=0.125,
                ylabel style={at={(axis description cs:-0.05,0.5)}, anchor=south, align=right},
                ytick={-0.1, -0.05, 0, 0.05, 0.1},
                yticklabel style={/pgf/number format/fixed, /pgf/number format/precision=2, text width=2.2em, align=right},
                ylabel={$x_w$ / m},
                axis background/.style={fill=white},
                axis x line*=bottom,
                axis y line*=left,
                xmajorgrids,
                ymajorgrids,
                xticklabels=\empty,
                title={Regulation task: \acrshort{LQR} run 3, \acrshort{NNC} run 2},
                title style={yshift=-1ex},
                legend style={legend columns=-1, font=\footnotesize, at={(0.005,1.05)}, anchor=north west, legend cell align=left, align=left, draw=white!15!black}
                ]
                \addplot [color=mycolor1]
                  table[]{Sigi_Run_LQR_film_Regulation_3-1.tsv};
                \addlegendentry{Reference}

                \addplot [color=mycolor3]
                  table[]{Sigi_Run_NN_film_Regulation_2-2.tsv};
                \addlegendentry{NNC}
                
                \addplot [color=mycolor2]
                  table[]{Sigi_Run_LQR_film_Regulation_3-2.tsv};
                \addlegendentry{LQR}
								
				\nextgroupplot
				[
				xmin=60,
                xmax=100,
                ymin=-0.125,
                ymax=0.125,
                ylabel style={at={(axis description cs:-0.05,0.5)}, anchor=south, align=right},                 
                yticklabel style={/pgf/number format/fixed, /pgf/number format/precision=2, text width=2.2em, align=right},
                ylabel={$\dot{x}_w$ / m/s},
                axis background/.style={fill=white},
                axis x line*=bottom,
                axis y line*=left,
                xmajorgrids,
                ymajorgrids,
                xticklabels=\empty,
                legend style={legend columns=-1, font=\footnotesize, at={(0.02,0.98)}, anchor=north west, legend cell align=left, align=left, draw=white!15!black}
                ]
                \addplot [color=mycolor1]
                  table[]{Sigi_Run_LQR_film_Regulation_3-3.tsv};

                \addplot [color=mycolor3]
                  table[]{Sigi_Run_NN_film_Regulation_2-4.tsv};
                
                \addplot [color=mycolor2]
                  table[]{Sigi_Run_LQR_film_Regulation_3-4.tsv};
                
                \nextgroupplot
				[
				xmin=60,
                xmax=100,
                ymin=-0.1,
                ymax=0.1,
                ylabel style={at={(axis description cs:-0.05,0.5)}, anchor=south, align=right},
                yticklabel style={/pgf/number format/fixed, /pgf/number format/precision=2, text width=2.2em, align=right},
                ylabel={$\theta$ / rad},
                axis background/.style={fill=white},
                axis x line*=bottom,
                axis y line*=left,
                xmajorgrids,
                ymajorgrids,
                xticklabels=\empty,
                legend style={legend columns=-1, font=\footnotesize, at={(0.02,0.98)}, anchor=north west, legend cell align=left, align=left, draw=white!15!black}
                ]
                \addplot [color=mycolor1]
                  table[]{Sigi_Run_LQR_film_Regulation_3-5.tsv};
                
                \addplot [color=mycolor3]
                  table[]{Sigi_Run_NN_film_Regulation_2-6.tsv};
                
                \addplot [color=mycolor2]
                  table[]{Sigi_Run_LQR_film_Regulation_3-6.tsv};
                
                \nextgroupplot
				[
				xmin=60,
                xmax=100,
                ymin=-2.5,
                ymax=2.5,
                ylabel style={at={(axis description cs:-0.05,0.5)}, anchor=south, align=right},
                yticklabel style={/pgf/number format/fixed, /pgf/number format/precision=2, text width=2.2em, align=right},
                ylabel={$\dot{\theta}$ / rad/s},
                axis background/.style={fill=white},
                axis x line*=bottom,
                axis y line*=left,
                xmajorgrids,
                ymajorgrids,
                xticklabels=\empty,
                legend style={legend columns=-1, font=\footnotesize, at={(0.02,0.98)}, anchor=north west, legend cell align=left, align=left, draw=white!15!black}
                ]
                \addplot [color=mycolor1]
                  table[]{Sigi_Run_LQR_film_Regulation_3-7.tsv};
                
                \addplot [color=mycolor3]
                  table[]{Sigi_Run_NN_film_Regulation_2-8.tsv};
                
                \addplot [color=mycolor2]
                  table[]{Sigi_Run_LQR_film_Regulation_3-8.tsv};
                
                \nextgroupplot
				[
				xmin=60,
                xmax=100,
                xlabel style={font=\color{white!15!black}},
                xlabel={$t$ / s},
                ymin=-2,
                ymax=2,
                ylabel style={at={(axis description cs:-0.05,0.5)}, anchor=south, align=right},
                yticklabel style={/pgf/number format/fixed, /pgf/number format/precision=2, text width=2.2em, align=right},
                ylabel={$u$ / V},
                axis background/.style={fill=white},
                xmajorgrids,
                ymajorgrids,
                legend style={legend columns=-1, font=\footnotesize, at={(0.02,0.98)}, anchor=north west, legend cell align=left, align=left, draw=white!15!black}
                ]

                \addplot [color=mycolor3]
                  table[]{Sigi_Run_NN_film_Regulation_2-9.tsv};
                
                \addplot [color=mycolor2]
                  table[]{Sigi_Run_LQR_film_Regulation_3-9.tsv};

			\end{groupplot}
		\end{tikzpicture}%
	\end{center}
		\caption{A comparison of the \acrshort{LQR} and \acrshort{NNC}  runs with the median \acrshort{RMSE} value in $x_w$ in the regulation task.}
	    \label{fig:OriginTracking}
\end{figure*}

\begin{figure*}[t]
	\begin{center}
		\begin{tikzpicture}
			\begin{groupplot}
				[
				group style={group name=MovingReferenceTrackingTrackingGroupPlot, group size=1 by 5, vertical sep=0.2cm, horizontal sep=1cm
				},
				width=18cm, height=2.8cm,
				]
				\nextgroupplot
				[
				xmin=60,
                xmax=100,
                ymin=-0.15,
                ymax=1.15,
                ylabel style={at={(axis description cs:-0.05,0.5)}, anchor=south, align=right},
                yticklabel style={/pgf/number format/fixed, /pgf/number format/precision=2, text width=2.2em, align=right},
                ylabel={$x_w$ / m},
                axis background/.style={fill=white},
                axis x line*=bottom,
                axis y line*=left,
                xmajorgrids,
                ymajorgrids,
                xticklabels=\empty,
                title={Reference-tracking task: \acrshort{LQR} run 5, \acrshort{NNC} run 2},
                title style={yshift=-1ex},
                legend style={legend columns=-1, font=\footnotesize, at={(0.02,0.98)}, anchor=north west, legend cell align=left, align=left, draw=white!15!black}
                ]
                \addplot [color=mycolor1]
                  table[]{Sigi_Run_LQR_film_Reference_5-1.tsv};
                \addlegendentry{Reference}

                \addplot [color=mycolor3]
                  table[]{Sigi_Run_NN_film_Reference_2-2.tsv};
                \addlegendentry{NNC}
                
                \addplot [color=mycolor2]
                  table[]{Sigi_Run_LQR_film_Reference_5-2.tsv};
                \addlegendentry{LQR}
								
				\nextgroupplot
				[
				xmin=60,
                xmax=100,
                ymin=-0.2,
                ymax=0.2,
                ylabel style={at={(axis description cs:-0.05,0.5)}, anchor=south, align=right},
                yticklabel style={/pgf/number format/fixed, /pgf/number format/precision=2, text width=2.2em, align=right},
                ylabel={$\dot{x}_w$ / m/s},
                axis background/.style={fill=white},
                axis x line*=bottom,
                axis y line*=left,
                xmajorgrids,
                ymajorgrids,
                xticklabels=\empty,
                legend style={legend columns=-1, font=\footnotesize, at={(0.02,0.98)}, anchor=north west, legend cell align=left, align=left, draw=white!15!black}
                ]
                \addplot [color=mycolor1]
                  table[]{Sigi_Run_LQR_film_Reference_5-3.tsv};

                \addplot [color=mycolor3]
                  table[]{Sigi_Run_NN_film_Reference_2-4.tsv};
                
                \addplot [color=mycolor2]
                  table[]{Sigi_Run_LQR_film_Reference_5-4.tsv};
                
                \nextgroupplot
				[
				xmin=60,
                xmax=100,
                ymin=-0.15,
                ymax=0.15,
                ylabel style={at={(axis description cs:-0.05,0.5)}, anchor=south, align=right},
                yticklabel style={/pgf/number format/fixed, /pgf/number format/precision=2, text width=2.2em, align=right},
                ylabel={$\theta$ / rad},
                axis background/.style={fill=white},
                axis x line*=bottom,
                axis y line*=left,
                xmajorgrids,
                ymajorgrids,
                xticklabels=\empty,
                legend style={legend columns=-1, font=\footnotesize, at={(0.02,0.98)}, anchor=north west, legend cell align=left, align=left, draw=white!15!black}
                ]
                \addplot [color=mycolor1]
                  table[]{Sigi_Run_LQR_film_Reference_5-5.tsv};

                \addplot [color=mycolor3]
                  table[]{Sigi_Run_NN_film_Reference_2-6.tsv};
                
                \addplot [color=mycolor2]
                  table[]{Sigi_Run_LQR_film_Reference_5-6.tsv};
                
                \nextgroupplot
				[
				xmin=60,
                xmax=100,
                ymin=-2.5,
                ymax=2.5,
                ylabel style={at={(axis description cs:-0.05,0.5)}, anchor=south, align=right},
                yticklabel style={/pgf/number format/fixed, /pgf/number format/precision=2, text width=2.2em, align=right},
                ylabel={$\dot{\theta}$ / rad/s},
                axis background/.style={fill=white},
                axis x line*=bottom,
                axis y line*=left,
                xmajorgrids,
                ymajorgrids,
                xticklabels=\empty,
                legend style={legend columns=-1, font=\footnotesize, at={(0.02,0.98)}, anchor=north west, legend cell align=left, align=left, draw=white!15!black}
                ]
                \addplot [color=mycolor1]
                  table[]{Sigi_Run_LQR_film_Reference_5-7.tsv};

                \addplot [color=mycolor3]
                  table[]{Sigi_Run_NN_film_Reference_2-8.tsv};

                \addplot [color=mycolor2]
                  table[]{Sigi_Run_LQR_film_Reference_5-8.tsv};

                \nextgroupplot
				[
				xmin=60,
                xmax=100,
                xlabel style={font=\color{white!15!black}},
                xlabel={$t$ / s},
                ymin=-2,
                ymax=2,
                ylabel style={at={(axis description cs:-0.05,0.5)}, anchor=south, align=right},
                yticklabel style={/pgf/number format/fixed, /pgf/number format/precision=2, text width=2.2em, align=right},
                ylabel={$u$ / V},
                axis background/.style={fill=white},
                xmajorgrids,
                ymajorgrids,
                legend style={legend columns=-1, font=\footnotesize, at={(0.02,0.98)}, anchor=north west, legend cell align=left, align=left, draw=white!15!black}
                ]

                \addplot [color=mycolor3]
                  table[]{Sigi_Run_NN_film_Reference_2-9.tsv};
                
                \addplot [color=mycolor2]
                  table[]{Sigi_Run_LQR_film_Reference_5-9.tsv};
			\end{groupplot}
		\end{tikzpicture}%
	\end{center}
	\caption{A comparison of the \acrshort{LQR} and \acrshort{NNC}  runs with the median \acrshort{RMSE} value in $x_w$ in the reference-tracking task.}
	    \label{fig:MovingReferenceTracking}
\end{figure*}
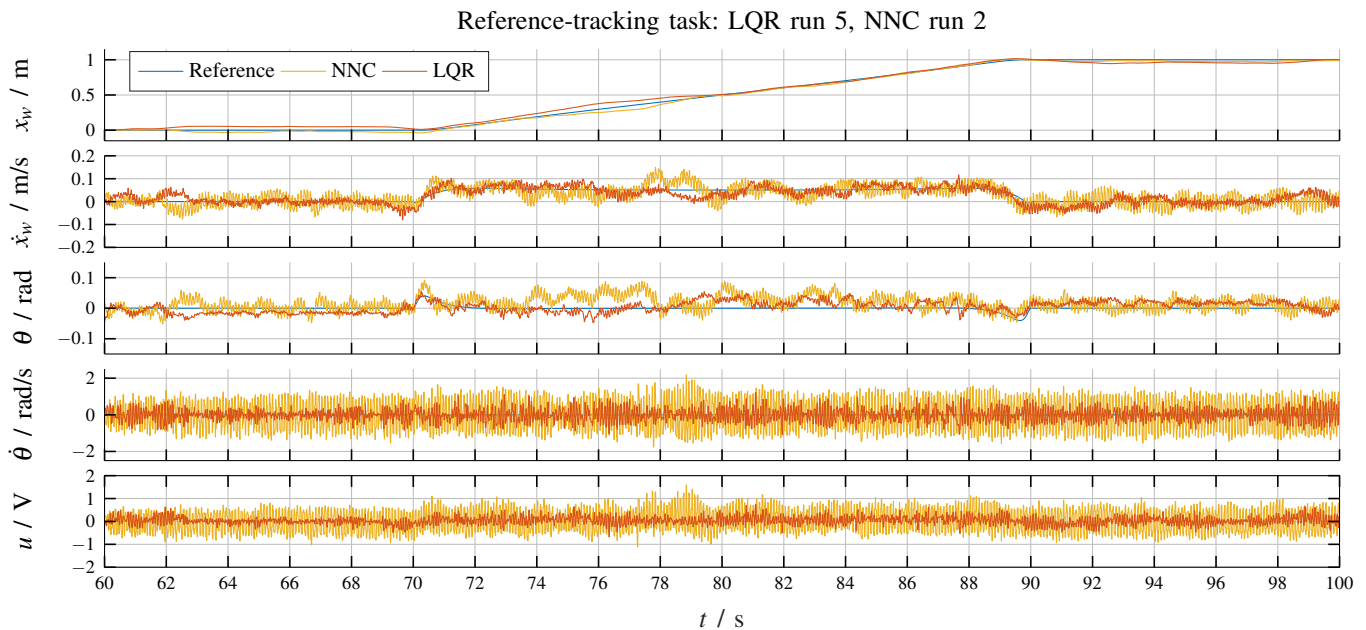

Each control task is run five times for both the \acrshort{LQR} and \acrshort{NNC}.\footnote{Video documentation of the corresponding experiments is available at https://www.youtube.com/@GDLab-b4g}
The average of the \acrfull{RMSE} and maximum of the \acrfull{MAE} over the five runs of all states $x_w$, $\dot{x}_w$, $\theta$, $\dot{\theta}$ as well as quantities $x_{\textrm{CoG}}$, $\dot{x}_{\textrm{CoG}}$ are reported and used as a metric of each controller's performance. These values are reported in \cref{tab:RegulationControlResults} and \cref{tab:ReferenceTrajectoryControlResults} for the regulation and reference-tracking task, respectively. In addition, for both the regulation and reference-tracking tasks,
the runs with the median \acrshort{RMSE} value in $x_w$ are displayed in \cref{fig:OriginTracking,fig:MovingReferenceTracking}, respectively. 

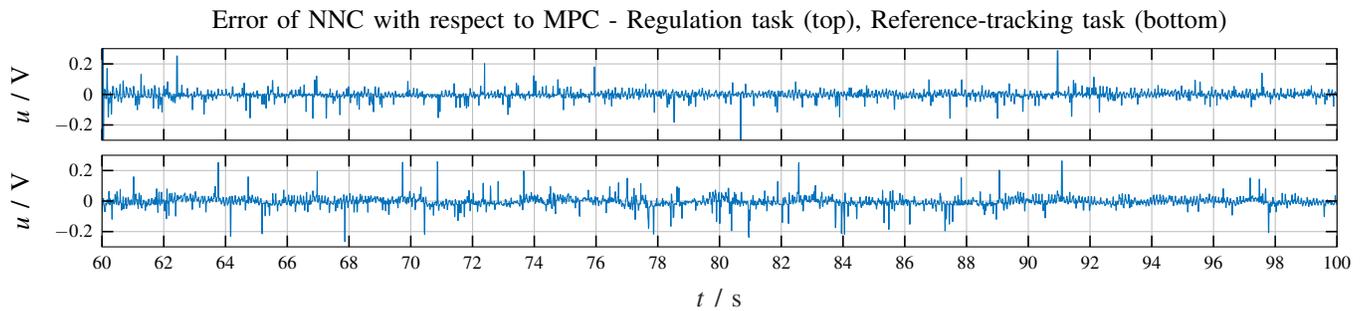
\begin{figure*}[tb]
	\begin{center}
		\begin{tikzpicture}      
			\begin{groupplot}
				[
				group style={group name=NNC2MPCerrorGroupPlot, group size=1 by 2, vertical sep=0.2cm, horizontal sep=1cm
				},
				width=18cm, height=2.8cm,
				]

                \nextgroupplot
				[
				xmin=60,
                xmax=100,
                xlabel style={font=\color{white!15!black}},
                ymin=-0.3,
                ymax=0.3,
                ylabel style={at={(axis description cs:-0.05,0.5)}, anchor=south, align=right},                 yticklabel style={/pgf/number format/fixed, /pgf/number format/precision=2, text width=2.2em, align=right},
                ylabel={$u$ / V},
                axis background/.style={fill=white},
                xmajorgrids,
                ymajorgrids,
                xticklabels=\empty,
                title={Error of \acrshort{NNC} with respect to \acrshort{MPC} - Regulation task (top), Reference-tracking task (bottom)},
                title style={ yshift=-1ex},
                legend style={legend columns=-1, font=\footnotesize, at={(0.02,0.98)}, anchor=north west, legend cell align=left, align=left, draw=white!15!black}
                ]

                \addplot [color=mycolor1]
                  table[]{Sigi_Run_LQR_film_Regulation_3_NN_film_Regulation_2_NNC_MPC_comp-1.tsv};

                \nextgroupplot
				[
				xmin=60,
                xmax=100,
                xlabel style={font=\color{white!15!black}},
                xlabel={$t$ / s},
                ymin=-0.3,
                ymax=0.3,
                ylabel style={at={(axis description cs:-0.05,0.5)}, anchor=south, align=right},                 yticklabel style={/pgf/number format/fixed, /pgf/number format/precision=2, text width=2.2em, align=right},
                ylabel={$u$ / V},
                axis background/.style={fill=white},
                xmajorgrids,
                ymajorgrids,
                legend style={legend columns=-1, font=\footnotesize, at={(0.02,0.98)}, anchor=north west, legend cell align=left, align=left, draw=white!15!black}
                ]

                \addplot [color=mycolor1]
                  table[]{Sigi_Run_LQR_film_Reference_5_NN_film_Reference_2_NNC_MPC_comp-1.tsv};
                
			\end{groupplot}   
		\end{tikzpicture}%
	\end{center}
        \caption{A comparison of the recorded control output of the \acrshort{NNC} and the \acrshort{MPC} control output computed at every recorded state \textit{a posteriori} for the runs shown in \cref{fig:OriginTracking,fig:MovingReferenceTracking}.}
	    \label{fig:NNC2MPCerror}
\end{figure*}

Using the recorded state trajectory, the theoretical input of the \acrshort{MPC} at every recorded state is calculated \textit{a posteriori} and compared to the recorded output value of the \acrshort{NNC}. The error between the output of these controllers, which was estimated and used as a design parameter in \cref{sec:ControllerSynthesis_RobustTubeMPC}, is displayed in \cref{fig:NNC2MPCerror} for the runs shown in \cref{fig:OriginTracking,fig:MovingReferenceTracking}.

\begin{table}[t]
\centering
\begin{tabular}{l|@{}c@{}c|@{}c@{}c|l}
\hline
\multirow{2}{*}{\textbf{Qty.}} 
& \multicolumn{2}{c|}{\textbf{Avg. \acrshort{RMSE}}} 
& \multicolumn{2}{c|}{\textbf{Max. \acrshort{MAE}}} 
& \multirow{2}{*}{\textbf{Unit}} \\ \cline{2-5} 
& \multicolumn{1}{l}{\: \textbf{LQR}\phantom{$a^b$}} 
& \textbf{\: \acrshort{NNC} \:} 
& \multicolumn{1}{l}{\: \textbf{LQR} \:} 
& \textbf{\: \acrshort{NNC} \:} 
& \\ \hline
$x_w$ & $\num[scientific-notation=true,round-precision=3]{2.788E-02}$ & $\mathbf{\num[output-exponent-marker={\bm{e}},scientific-notation=true,round-precision=3]{1.682E-02}}$ & $\num[scientific-notation=true,round-precision=3]{8.497E-02}$ & $\mathbf{\num[output-exponent-marker={\bm{e}},scientific-notation=true,round-precision=3]{4.558E-02}}$ & $\si{\metre}$ \\
$\dot{x}_w$ & $\mathbf{\num[output-exponent-marker={\bm{e}},scientific-notation=true,round-precision=3]{2.549E-02}}$ & $\num[scientific-notation=true,round-precision=3]{4.098E-02}$ & $\mathbf{\num[output-exponent-marker={\bm{e}},scientific-notation=true,round-precision=3]{8.842E-02}}$ & $\num[scientific-notation=true,round-precision=3]{1.199E-01}$ & $\si{\metre\per\second}$ \\
$\theta$ & $\mathbf{\num[output-exponent-marker={\bm{e}},scientific-notation=true,round-precision=3]{1.711E-02}}$ & $\num[scientific-notation=true,round-precision=3]{2.421E-02}$ & $\num[scientific-notation=true,round-precision=3]{8.833E-02}$ & $\mathbf{\num[output-exponent-marker={\bm{e}},scientific-notation=true,round-precision=3]{8.538E-02}}$ & $\si{\radian}$ \\
$\dot{\theta}$ & $\mathbf{\num[output-exponent-marker={\bm{e}},scientific-notation=true,round-precision=3]{5.596E-01}}$ & $\num[scientific-notation=true,round-precision=3]{1.030E+00}e\!-\!0$ & $\mathbf{\num[output-exponent-marker={\bm{e}},scientific-notation=true,round-precision=3]{1.584E+00}e\!-\!0}$ & $\num[scientific-notation=true,round-precision=3]{2.103E+00}e\!-\!0$ & $\si{\radian\per\second}$ \\
\end{tabular}
\caption{An overview of the average \acrshort{RMSE} and the maximum \acrshort{MAE} of the system states and linearized \acrshort{CoG} position/velocity over five runs of the regulation task (lowest in bold).}
\label{tab:RegulationControlResults}
\end{table}

\begin{table}[t]
\centering
\begin{tabular}{l|@{}c@{}c|@{}c@{}c|l}
\hline
\multirow{2}{*}{\textbf{Qty.}} 
& \multicolumn{2}{c|}{\textbf{Avg. \acrshort{RMSE}}} 
& \multicolumn{2}{c|}{\textbf{Max. \acrshort{MAE}}} 
& \multirow{2}{*}{\textbf{Unit}} \\ \cline{2-5} 
& \multicolumn{1}{l}{\: \textbf{LQR}\phantom{$a^b$}} 
& \textbf{\: \acrshort{NNC} \:} 
& \multicolumn{1}{l}{\: \textbf{LQR} \:} 
& \textbf{\: \acrshort{NNC} \:} 
& \\ \hline
$x_w$ 
& $\num[scientific-notation=true,round-precision=3]{4.018E-02}$ 
& $\mathbf{\num[output-exponent-marker={\bm{e}},scientific-notation=true,round-precision=3]{2.482E-02}}$ 
& $\num[scientific-notation=true,round-precision=3]{1.384E-01}$ 
& $\mathbf{\num[output-exponent-marker={\bm{e}},scientific-notation=true,round-precision=3]{8.401E-02}}$ 
& $\si{\metre}$ \\
$\dot{x}_w$ 
& $\mathbf{\num[output-exponent-marker={\bm{e}},scientific-notation=true,round-precision=3]{2.130E-02}}$ 
& $\num[scientific-notation=true,round-precision=3]{3.236E-02}$ 
& $\mathbf{\num[output-exponent-marker={\bm{e}},scientific-notation=true,round-precision=3]{8.098E-02}}$ 
& $\num[scientific-notation=true,round-precision=3]{2.670E-01}$ 
& $\si{\metre\per\second}$ \\
$\theta$ 
& $\mathbf{\num[output-exponent-marker={\bm{e}},scientific-notation=true,round-precision=3]{1.811E-02}}$ 
& $\num[scientific-notation=true,round-precision=3]{2.615E-02}$ 
& $\mathbf{\num[output-exponent-marker={\bm{e}},scientific-notation=true,round-precision=3]{7.043E-02}}$ 
& $\num[scientific-notation=true,round-precision=3]{3.263E-01}$ 
& $\si{\radian}$ \\
$\dot{\theta}$ 
& $\mathbf{\num[output-exponent-marker={\bm{e}},scientific-notation=true,round-precision=3]{3.981E-01}}$ 
& $\num[scientific-notation=true,round-precision=3]{8.233E-01}$ 
& $\mathbf{\num[output-exponent-marker={\bm{e}},scientific-notation=true,round-precision=3]{1.585E+00}e\!-\!0}$ 
& $\num[scientific-notation=true,round-precision=3]{4.493E+00}e\!-\!0$ 
& $\si{\radian\per\second}$ \\
\end{tabular}
\caption{An overview of the average \acrshort{RMSE} and the maximum \acrshort{MAE} of the system states and linearized \acrshort{CoG} position/velocity over five runs of the reference-tracking task (lowest in bold).}
\label{tab:ReferenceTrajectoryControlResults}
\end{table}

The empirical performance results demonstrate the improved performance of the \acrshort{NNC} over the baseline \acrshort{LQR}. The \acrshort{RMSE} values of \cref{tab:RegulationControlResults,tab:ReferenceTrajectoryControlResults} indicate that the \acrshort{NNC} tracks the $x_w$ and $x_{\textrm{CoG}}$ quantities significantly better than the \acrshort{LQR} in both the regulation and reference-tracking tasks. However, the \acrshort{RMSE} of the \acrshort{NNC} is generally larger for all other quantities. This can be explained by the system's unstable and non-minimum phase behavior, the \acrshort{NNC}'s nonlinear behavior near the boundary of the feasible set of the underlying \acrshort{MPC}, and the relatively aggressive value of $\rho$ in \cref{tab:TubeMPCParameters}. 

Furthermore, via an examination of the maximum \acrshort{MAE} of both controllers in the regulation and reference-tracking tasks, it is clear that the \acrshort{LQR} does not satisfy the constraints of \cref{tab:TubeMPCParameters} used to define the \acrshort{MPC}. On the other hand, as suggested by the comparison in \cref{sec:Results_QualitativeComparison}, the \acrshort{NNC} does approximately satisfy these constraints, thereby improving the control performance. An example of this nonlinear behavior, which has been learnt from the robust, tube-based \acrshort{MPC}, can be seen at $t \in [ \qty[scientific-notation=false,round-precision=2]{77}{\second}, \, \qty[scientific-notation=false,round-precision=2]{79}{\second}]$ in \cref{fig:MovingReferenceTracking}. The large negative control input prevents the position error accumulating beyond the prescribed state constraints. It should be noted that the margin by which the \acrshort{NNC} violates the constraints can be partially attributed to the backlash present in the system.

Moreover, \cref{fig:NNC2MPCerror} indicates that the \acrshort{NNC} approximates the \acrshort{MPC} law well, with the approximation error generally remaining below the value of $\qty[scientific-notation=false,round-precision=2]{0.075}{\volt}$ assumed in \cref{sec:ControllerSynthesis_RobustTubeMPC}, thereby supporting the validity of this design parameter.

Overall, these results demonstrate the effectiveness of the generalizable synthesis procedure presented in \cref{sec:ControllerSynthesis_NNC} and motivate the use of the \acrshort{SOS}-based stability verification procedure of \cref{sec:StabilityVerification} to obtain \acrshortpl{NNC} with known stability properties capable of outperforming simple model-based controllers, e.g. an \acrshort{LQR}, on systems with limited compute capabilities.

\section{Conclusion}
\label{sec:Conclusion}
This work presents a generalizable synthesis procedure for an \acrshort{NNC} to imitate a robust, tube-based MPC and is the first to establish the feasibility and practical value of an \acrshort{SOS}-based stability verification procedure for applied control problems by verifying local stability properties of an \acrshort{NNC} synthesized to control the Sigi platform, a two-wheeled inverted pendulum for which a control-oriented model and state estimator are available. Using the generalizable procedure a \acrshort{NNC} is synthesized to imitate a robust, tube-based \acrshort{MPC} controller which itself is computationally too expensive to run in real time on the available embedded hardware. By means of the \acrshort{SOS}-based stability verification procedure of \cref{sec:StabilityVerification}, the two-wheeled inverted pendulum is proven to be \acrshort{LAS} under this \acrshort{NNC}, and a relevant inner estimate of the \acrshort{ROA} is obtained. Empirical performance data verifies the improved performance of the \acrshort{NNC} over a baseline \acrshort{LQR} in both regulation and reference-tracking tasks. Taken together, these results illustrate the value of the generalizable synthesis procedure for obtaining stabilizing \acrshortpl{NNC} capable of approximating complex control strategies while maintaining low computational requirements. In addition, they establish the practical value of the the \acrshort{SOS}-based stability verification procedure used in this work for real-world and/or safety-critical control problems.

Future work will focus on a variety of aspects of the \acrshort{SOS}-based stability verification procedure to further improve its practical value. This includes improved problem formulations \cite[Section IV.C]{mybibfile:Detailleur2025}, methods to reduce the computational complexity, potentially allowing larger \acrshortpl{NNC} to be examined, and extending the framework to allow the analysis of \acrshortpl{NNC} utilizing different architectures and activation functions\cite[Section III]{mybibfile:Detailleur2025}. Furthermore, the possibility of controller synthesis, in which the SOS stability certificate is constructed during the training process, should be explored.

\section{Acknowledgment}
We acknowledge Ilyas Seckin and Joschua Wüthrich for fruitful discussions and meaningful inputs to this work.

\section*{References}
\bibliographystyle{IEEEtranBST/IEEEtran}
\renewcommand{\section}[2]{}%
\bibliography{IEEEtranBST/IEEEabrv,mybibfile}

\vskip -2\baselineskip plus -1fil
\begin{IEEEbiographynophoto}{Alvaro Detailleur}
received the M.Sc. degree in robotics, systems and control from ETH Zurich, Zurich, Switzerland in 2024, focusing on system modeling and model-based control design. 

He has completed industrial internships at Forze Hydrogen Racing and Mercedes-AMG High Performance Powertrains, focusing on the software and control of high performance automotive power units. His research interests include nonlinear control, neural-network-based controllers and optimization-based control design with a practical application.

Mr. Detailleur was awarded the Young Talent Development Prize by the Royal Dutch Academy of Sciences (KHMW).
\end{IEEEbiographynophoto}
\vskip -2\baselineskip plus -1fil
\begin{IEEEbiographynophoto}{Dalim Wahby} received the B.Sc. degree in industrial engineering and management from Karlsruhe Institute of Technology (KIT), Karlsruhe, Germany, in 2022, with a focus on energy technologies and natural language processing. Additionally, he received the Dipl. Ing. in electronics and embedded systems from Polytech Nice-Sophia, Sophia Antipolis, France in 2024, and the M.Sc. in ICT innovation from Royal Institute of Technology (KTH), Stockholm, Sweden in 2025, with a major in electrical engineering.

He has completed a research internship at CNRS, focusing on adaptive control and the stability analysis of neural-network-based controllers. Currently, he is pursuing the Ph.D. degree in automatic signal and image processing at i3S/CNRS in Sophia-Antipolis, under the supervision of Guillaume Ducard, focusing on the development of a framework for the design and the analysis of neural-network-based controllers.
\end{IEEEbiographynophoto}
\vskip -2\baselineskip plus -1fil
\begin{IEEEbiographynophoto}{Guillaume Ducard} (Senior Member, IEEE), received the M.Sc. degree in electrical engineering and the 
Doctoral degree focusing on flight control for unmanned aerial vehicles (UAVs) from ETH
Zurich, Zurich, Switzerland, in 2004 and 2007, respectively.

He completed his two-year Postdoctoral course in 2009 from ETH Zurich, focused
on flight control for UAVs. He is currently an Associate Professor with the
Universit{\'e} C\^{o}te d`Azur, France, and guest scientist with ETH Zurich. His research interests include nonlinear control, neural networks, estimation, and guidance mostly applied to UAVs.
\end{IEEEbiographynophoto}
\vskip -2\baselineskip plus -1fil
\begin{IEEEbiographynophoto}{Christopher H. Onder} received the Diploma in mechanical engineering and Doctoral
degree in Doctor of technical sciences from ETH Zurich, Zurich, Switzerland.

He is a Professor with the Institute for Dynamic Systems and Control, Department of
Mechanical Engineering and Process Control, ETH Zurich. He heads the Engine Systems
Laboratory, and has authored and co-authored numerous articles and a book on modeling and
control of engine systems. His research interests include engine systems modeling, control
and optimization with an emphasis on experimental validation, and industrial cooperation.

Prof. Dr. Onder was the recipient of the BMW scientific award, the ETH medal, the
Vincent Bendix award, the Crompton Lanchester Medal, and the Arch T. Colwell award.
Additionally, he was awarded the Watt d’Or, the energy efficiency price of the
Swiss Federal Office of Energy, on multiple occasions for his projects. 
\end{IEEEbiographynophoto}
\end{document}